\documentclass[a4paper,onefignum]{siamart220329}
\usepackage{lipsum}
\usepackage{amsfonts,amsmath,amssymb}
\usepackage{graphicx}
\usepackage{epstopdf}
\usepackage{algorithmic}
\usepackage{float,subfigure,tikz}
\usepackage{xcolor}
\usepackage{bm}
\usepackage{pgfplots}
\usepackage{amsopn}
\usetikzlibrary{spy}
\ifpdf
  \DeclareGraphicsExtensions{.eps,.pdf,.png,.jpg}
\else
  \DeclareGraphicsExtensions{.eps}
\fi

\newsiamremark{remark}{Remark}
\newsiamremark{hypothesis}{Hypothesis}
\crefname{hypothesis}{Hypothesis}{Hypotheses}
\newsiamthm{claim}{Claim}
\newsiamthm{example}{Example}

\headers{FW-BICs in a Cavity with a thin
  Waveguide Opening}{Jiaxin Zhou, Wangtao Lu and Ya Yan Lu}

\title{Existence of Friedrich-Wintgen Bound States in the
    Continuum: Cavity with a Thin
  Waveguide Opening}

  \author{
  Jiaxin Zhou\thanks{Liu Bie Ju Centre for Mathematical
    Sciences, City University of Hong Kong, Kowloon, Hong Kong, China
    (\email{jiaxzhou@cityu.edu.hk}).}    
  \and Wangtao Lu\thanks{School of Mathematical Sciences, Zhejiang
    University, Hangzhou 310027, China
    (\email{wangtaolu@zju.edu.cn}).}
  \and Ya Yan Lu\thanks{Department of Mathematics, City University of
    Hong Kong, Kowloon, Hong Kong, China
    (\email{mayylu@cityu.edu.hk}).}
}

\def\bi{{\bf i}}
\definecolor{orange}{rgb}{1,0.7,0}

\begin{document}

\maketitle

\begin{abstract}
    Bound states in the continuum (BICs) are localized states
    embedded within a continuum of propagating waves.
    Perturbations that disrupt BICs typically induce 
    ultra-strong resonances, a phenomenon enabling diverse
    applications in photonics.    
    This work investigates the existence of BICs in two-dimensional    
    electromagnetic cavities coupled to thin waveguides for H-polarized
    waves. Our focus is on Friedrich-Wintgen BICs (FW-BICs), which arise
    from destructive interference between two resonant modes and were
    identified numerically in rectangular cavities with waveguide
    openings by Lyapina et    
    al. [\emph{J. Fluid Mech.}, 780 (2015), pp. 370--387].
    Here, we rigorously  
    establish the existence of FW-BICs in a broader
    class of cavity geometries by introducing perturbations to
    the refractive index under regularity constraints.
    Assuming the waveguide width is sufficiently small,
    we show that BICs correspond to intersections  
    of two curves derived implicitly from the governing
    equations constructed via the mode-matching method.
    Crucially, we prove that such intersections are
    guaranteed, provided that
    two eigenvalues of the cavity intersect transversally
    and the associated eigenfunctions exhibit non-vanishing coupling
    to the radiation channel at the cavity-waveguide interface.
    Furthermore, our approach remains applicable for studying the
    emergence of FW-BICs under
    parameter-dependent boundary perturbations to
    the cavity.
\end{abstract}



\section{Introduction}
Bound states in the continuum (BICs), first proposed by von Neumann
and Wigner \cite{vonneumann29}, are trapped or guided modes that
coexist with
radiating waves capable of carrying energy away
\cite{hsu16,koshelev23,sadreev21}.
These states
have been observed in numerous wave systems governed by Schr$\ddot{\mathrm{o}}$dinger
equation \cite{friedrich85,stillingerherrick75}, Helmholtz equation
\cite{anne94,evans94,lyapina15,porter05,shipman03,shipman07},
Maxwell's equations \cite{hsuzhen13,jinyin19,yuanlu21}, etc.
Additionally, perturbing these wave 
systems slightly generally leads to resonances with ultra-high quality 
factor $(Q)$ which can be excited by incident waves,
causing local field enhancement and sharp features in transmission and
reflection spectra
\cite{zhen20,junshanstephenhai20,shipman12,shipman05}. Thus,
BICs can also be interpreted as resonances with purely real
frequencies. In recent years, BICs have found many
applications in photonics, including lasing
\cite{hwang21,kodigala17}, sensing  
\cite{yesilkoy2019}, nonlinear optics \cite{Koshelev19,yuan17,yuan20_2},
etc. Mathematical {analysis} on the properties of BICs remain
highly relevant and meaningful.

Friedrich-Wintgen BICs (FW-BICs) arise from    
destructive interference between resonant modes interacting
through a single radiation channel. Their mechanism was
first discovered by Friedrich and Wintgen in a system of three
one-dimensional (1D) Schr$\ddot{\mathrm{o}}$dinger equations
\cite{friedrich85}. A rigorous analysis of such BICs
within the Schr$\ddot{\mathrm{o}}$dinger system can be found in
\cite{yu25}.
Recently, FW-BICs have also 
been identified in 2D acoustic  
systems, particularly in rectangular cavities coupled to
semi-infinite waveguides \cite{lyapina15,maksimov15}. In these
studies, a mode-matching method was employed to construct an
infinite-dimensional linear system, where BICs emerge as non-trivial
solutions to the homogeneous problem. By neglecting evanescent
waves, the authors obtained an approximate
model that suggests BICs may
arise near multiple eigenvalues of the
rectangular cavity. Numerical experiments further demonstrate that
such BICs can be realized by introducing slight
perturbations to the cavity length. However, a rigorous mathematical
proof confirming the existence of these BICs remains elusive.

We consider a broader class of 2D cavities, each with a
semi-infinite waveguide attached to a segment of    
its boundary. The scattering matrix for wave systems in
such structures was previously studied by Adamyan
et. al. \cite{adamyan14}. Let $h$ denote the waveguide width
and $\mu$ the square of the free-space wavenumber.
As $h\to0$, the resonances ${\lambda_{\mathrm{res}}}$ of these open systems
converge to the eigenvalues $\lambda$ of the corresponding closed
cavities. The authors employ a two-step strategy, pioneered
by H. Poincar{\'e} \cite{poincare92}, to reveal the singular
term $(\overline{{\lambda_{\mathrm{res}}}}-\mu)/({\lambda_{\mathrm{res}}}-\mu)$
in the scattering matrix.
  
In this work, we focus on establishing sufficient conditions for the
existence of  
FW-BICs in the cavity-waveguide structures. Introducing an analytic
perturbation to the  
refractive index parameterized by $\delta$, and assuming $h\ll1$, 
we reduce the infinite-dimensional system derived from the
mode-matching method to two governing  
equations for $\delta$ and $\mu$ in small intervals. Each equation defines
an implicit function $\mu(\delta)$, and BICs are identified
at the intersections of these curves in the $(\delta,\mu)$-plane.
We demonstrate that such intersections are guaranteed to exist
provided that two cavity eigenvalues 
{intersect transversally} as $\delta$ varies, and  
the associated eigenfunctions exhibit non-vanishing coupling
to the radiation channel at the cavity-waveguide interface.
The main result is formalized in Theorem
\ref{thm:sec5:BIC:existence}. Our approach remains valid  
for demonstrating the existence of BICs under parameter-dependent
boundary perturbations, if the eigenvalues and eigenfunctions
exhibit continuous dependence on these perturbations. This
continuity condition is satisfied in specific scenarios (cf. Section
VII.6.5 in \cite{kato95}), including scaling transformations of
rectangular cavities as explored in \cite{lyapina15}. For additional
applications of analogous mode-matching techniques in resonance
analysis of subwavelength structures, we refer the readers to
\cite{linlu23,luwangzhou24,zhou21,zhou25}.

The structure of this paper is as follows. Section 2 is devoted to the
problem formulation and the three key conditions underlying the
analysis. The derivation of the governing infinite-dimensional linear
system is presented in Section 3. Section 4 details the reformulation
of this system into a real-valued form and its subsequent reduction to
a finite-dimensional system. A rigorous proof for the existence of
FW-BICs, which is the central result of this paper, is established in
Section 5. Numerical
simulations validating the theoretical findings are provided in
Section 6. The paper concludes in Section 7 with a discussion on the
extension of these results to boundary perturbations and other
generalizations of the approach.

\section{Problem formulation and key requirements}
\label{sec2:MainContribution}
As illustrated in Fig. \ref{fig:sec2:cavity-waveguide:axis}, our 
problem region
$\Omega$ is defined as $\Omega:=\Omega_{\mathrm{in}}\cup\Gamma_{h}\cup\Omega_{\mathrm{out}}$, where
$\Omega_{\mathrm{in}}$ represents the cavity region with a line segment on
its boundary, $\Omega_{\mathrm{out}}$ denotes
the waveguide region of width $h$ extended from the line segment, and
$\Gamma_{h}$ is the common boundary between these two regions. Letting
$\mathbf{o}$ denote the middle point of $\Gamma_{h}$, we establish a rectangular
coordinate system in $\mathbb{R}^{2}$ centered at
$\mathbf{o}$. Without loss of generality, we always fix
the point $\mathbf{o}$ on $\partial\Omega_{\mathrm{in}}$ as the middle point of
$\Gamma_{h}$ when $h$ is allowed to decrease in the subsequent sections.
Additionally, $\Omega_{B}$ denotes a rectangular region in
$\Omega_{\mathrm{in}}$ such that $\Gamma_{h}\subset\partial\Omega_{B}$.
\begin{figure}[htbp]
  \centering
    \begin{tikzpicture}
      \begin{scope}
  [spy using outlines={rectangle, blue, magnification=1.5,
    size=2cm, connect spies}]        
\draw [black,ultra thick] plot [smooth] coordinates {(5,-1) (4,-2) (3,-2) (2,-1.5)
  (1,-1.2) (0,0) (2,1) (3,1) (5,-0)};
\fill [lightgray] plot [smooth] coordinates { (3.3,-1) (2.1,-1.2)
  (1.1,0) (3.1,0.5) (3.6,0) (3.3,-1)};
\draw [ultra thick](5,0) -- (5,-0.25);
\draw [ultra thick](5,-0.25) -- (10,-0.25);
\draw [ultra thick](5,-0.75) -- (5,-1);
\draw [ultra thick](5,-0.75) -- (10,-0.75);
\draw [dashed,ultra thick] (5,0) -- (5,-1);
\draw (3.8,-1.5) node[]{$\Omega_{\mathrm{in}}$};
\draw (2.5,-0.3) node[]{$n(\mathbf{x})$};
\draw (8,-0.5) node[]{$\Omega_{\mathrm{out}}$};
\draw [<->,thick](9,-0.25) -- (9,-0.75) node [right,pos=0.5] {$h$};
\spy on (5,-0.5) in node [right] at (6.5,1.2);
\end{scope}
\draw (5.15,-0.62) node[red]{$\mathbf{o}$};
\draw [->,thick,red,dashed](5,-0.5) -- (6.5,-0.5) node [right,pos=1] {$x_{1}$};
\draw [->,thick,red,dashed](5,-0.5) -- (5,1) node [above,pos=1] {$x_{2}$};
\draw [dashed,ultra thick,brown](7.5,1.8) -- (7.5,0.6) -- (6.8,0.6) -- (6.8,1.8) -- (7.5,1.8);
\draw (4.7,-0.5) node[]{${\Gamma_{h}}$};
\draw (7.2,1.2) node[brown]{${\Omega_{B}}$};
\end{tikzpicture}        
    \label{fig:sec2:cavity-waveguide:axis}
    \caption{A cavity with a thin waveguide opening is
      represented as      
      $\Omega:=\Omega_{\mathrm{in}}\cup\Gamma_{h}\cup\Omega_{\mathrm{out}}$, where
$\Omega_{\mathrm{in}}$ represents the cavity region featuring a line segment on
its boundary, $\Omega_{\mathrm{out}}$ denotes
the waveguide region of width $h$ extended from the line segment vertically, and
$\Gamma_{h}$ is the common boundary between these two
regions. Additionally, $n(\mathbf{x})$ denotes the refractive index in
$\Omega$, and $\Omega_{B}$ denotes a rectangular region in $\Omega_{\mathrm{in}}$ that $\Gamma_{h}\subset\partial\Omega_{B}$.}
\end{figure}

We also assume $\Omega_{\mathrm{in}}$ is Lipschitz and use
$n(\mathbf{x})\in{L^{\infty}(\Omega)}$ to denote the
refractive index in $\Omega$ satisfying
\begin{equation}
  \label{eq:sec2:nsetup}
  \left\{\begin{aligned}
      &n_{0}\le{n}(\mathbf{x})\le{n_{1}},\ &&\text{if}\
        \mathbf{x}\in\Omega_{\mathrm{in}};\\
      &n(\mathbf{x})=1,\ &&\text{if}\ \mathbf{x}\in\Omega_{\mathrm{out}}\cup\Omega_{B},\\
    \end{aligned}\right.
\end{equation}
where $n_{0},n_{1}\in\mathbb{R}^{+}$. Letting $\mathcal{P}$ denote the elliptic operator
\begin{equation}
  \label{eq:sec2:operatorP}
   \mathcal{P}=-\nabla\cdot\left(\frac{1}{n^{2}}\nabla\right),
 \end{equation}
 we consider H-polarized electromagnetic waves in $\Omega$
   satisfying the Perfect Electric Conductor (PEC) boundary condition:
\begin{align}
  -\mathcal{P}{u}+\mu{u}&=0\ \mathrm{in}\
  \Omega,\ \label{eq:sec2:goveq}\\
  \partial_{\bm{\nu}}u&=0\ \mathrm{on}\ \partial\Omega,\label{eq:sec2:bc}
\end{align}
where $u$ is the only non-zero component of the magnetic field,
$\mu$ denotes the square of free space wavenumber and $\bm{\nu}$ is
the outward unit normal on $\partial\Omega$. The corresponding eigenvalue problem
in $\Omega_{\mathrm{in}}$ is given by 
\begin{align}
  -\mathcal{P}{\psi}+\lambda{\psi}&=0\ \mathrm{in}\
  \Omega_{\mathrm{in}},\ \label{eq:sec2:eigen:goveq}\\
  \partial_{\bm{\nu}}\psi&=0\ \mathrm{on}\ \partial\Omega_{\mathrm{in}},\label{eq:sec2:eigen:bc}
\end{align}
where $(\lambda,\psi)$ denotes an eigenpair.

Let the refractive index $n(\cdot,\delta)$ depend on another parameter
$\delta$. When $\delta=0$, we assume the problem
\eqref{eq:sec2:eigen:goveq}--\eqref{eq:sec2:eigen:bc} admits a multiple
eigenvalue of
multiplicity 2 at $\lambda_{M-2}=\lambda_{M-1}$ with corresponding eigenfunctions
$\psi_{M-2}$ and $\psi_{M-1}$ for some integer $M\ge3$. Here, the
subscripts denote the indices of the eigenvalues in ascending order.
We investigate the existence of Friedrich-Wintgen BICs satisfying
\eqref{eq:sec2:goveq}--\eqref{eq:sec2:bc} for $\delta\approx0$ and
  $\mu\approx\lambda_{M-1}$ when $h\ll1$, under the following conditions:
\begin{itemize}
\item[A.1.] The refractive index $n(\mathbf{\cdot},\delta)$ depends
  analytically on $\delta\in[-\delta_{0},\delta_{0}]$ where $\delta_{0}>0$ and
  satisfies \eqref{eq:sec2:nsetup} for each $\delta$.
\item[A.2.] When $\delta=0$, $\psi_{M-2}(\mathbf{o})\ne0$ and $\psi_{M-1}(\mathbf{o})\ne0$.
\item[A.3.] 
  At $\delta=0$, the eigenvalue functions
    $\lambda_{M-2}(\delta)$ and $\lambda_{M-1}(\delta)$ {intersect transversally}, resulting in an
    eigenvalue of multiplicity 2.
  \end{itemize}
A brief discussion of the necessity of these conditions can be found in Remark \ref{rmk:sec5:necessity}.  

\section{The mode-matching method}
In the $\Omega_{\mathrm{in}}$ region, we expand the waves by the
eigenfunctions of
\eqref{eq:sec2:eigen:goveq}--\eqref{eq:sec2:eigen:bc}, while in the
$\Omega_{\mathrm{out}}$ region, waves are expanded by outgoing
or exponentially decaying waveguide modes. Matching these
expansions on $\Gamma_{h}$ through the continuity of traces
and the second Green identity leads to an
infinite-dimensional linear system. And a non-trivial
solution of the homogeneous problem corresponds to a BIC.

\subsection{Eigenfunction expansion in the cavity}
For each $\delta$, let
\begin{equation}
  \label{eq:sec3:eigen}
    \lambda_{0}(\delta),\ \lambda_{1}(\delta),\ldots\ \text{and}\ \psi_{0}(\cdot,\delta),\ \psi_{1}(\cdot,\delta),\ldots    
  \end{equation}
denote the eigenvalues and corresponding normalized eigenfunctions of
$\mathcal{P}$ in $\Omega_{\mathrm{in}}$
satisfying \eqref{eq:sec2:eigen:goveq}--\eqref{eq:sec2:eigen:bc}.
\begin{lemma}
  \label{lem:sec3:eigen:continuity}
  Under Condition A.1, we can choose a set of normalized eigenfunctions
  $\{\psi_{m}(\cdot,\delta)\}$ such that the mapping $\delta\to\psi_{m}(\cdot,\delta)$ is continuous from  
  $[-\delta_{0},\delta_{0}]$ to $H^{1}(\Omega_{\mathrm{in}})$ for each $m\in\mathbb{N}$, i.e.,
  \begin{equation}
    \label{eq:sec3:H1converge}
    \|\psi_{m}(\cdot,\delta_{a})-\psi_{m}(\cdot,\delta_{b})\|_{H^{1}(\Omega_{\mathrm{in}})}\to0\
    \text{as}\ \delta_{a}\to\delta_{b}\ \text{for}\ \delta_{a},\delta_{b}\in[-\delta_{0},\delta_{0}].
  \end{equation}  
  Furthermore, the associated eigenvalue satisfies  
  \begin{equation}
    \label{eq:sec3:continuity:eigenvalue}
    \left|\frac{1}{\lambda_{m}(\delta_{a})+1}-\frac{1}{\lambda_{m}(\delta_{b})+1}\right|\le{e^{C_{0}|\delta_{a}-\delta_{b}|}-1},\
    \text{for}\ \delta_{a},\delta_{b}\in[-\delta_{0},\delta_{0}],
  \end{equation}
  where $C_{0}>0$ is a constant independent of $m$.
  \begin{proof}
    Let $\Phi_{\delta}(\cdot,\cdot)$ denote the corresponding sesquilinear form of
    \eqref{eq:sec2:operatorP} for each $\delta$:
    \begin{equation}
      \label{eq:sec3:sesform}
      \Phi_{\delta}(u,v):=(\frac{1}{n^{2}(\cdot,\delta)}\nabla{u},\nabla{v})_{\Omega_{\mathrm{in}}},\ u,{v}\in{H}^{1}(\Omega_{\mathrm{in}}).
    \end{equation}
    Finding eigenpairs $(\lambda,\psi)$ of
    \eqref{eq:sec2:eigen:goveq}--\eqref{eq:sec2:eigen:bc} is equivalent
    to solve    
    \begin{equation}
      \label{eq:sec3:eigen:sesform}
      \Phi_{\delta}(\psi,v)=\lambda(\psi,v)_{\Omega_{\mathrm{in}}},\ \text{for\ each}\ {v}\in{H}^{1}(\Omega_{\mathrm{in}}).
    \end{equation}
    Since $n^{2}(\cdot,\delta)$ varies analytically with respect to $\delta$ and
    $H^{1}(\Omega_{\mathrm{in}})$ is dense in $L^{2}(\Omega_{\mathrm{in}})$,
    $\Phi_{\delta}$ is a self-adjoint holomorphic family of forms of type (a)
    (cf. Sections VII.4.2 and VII.4.8 in \cite{kato95}). Hence,
    $\psi_{m}(\cdot,\delta)$ can be chosen as a continuous mapping from
    $[-\delta_{0},\delta_{0}]$ to    
    $L^{2}(\Omega_{\mathrm{in}})$, and its eigenvalue $\lambda_{m}(\delta)$ satisfies    
    \eqref{eq:sec3:continuity:eigenvalue} by using
    the results in Section VII.4.7 of \cite{kato95}.
    { To prove \eqref{eq:sec3:H1converge}, it suffices to
      show for each $m\in\mathbb{N}$ that
      \begin{equation}
        \label{eq:sec3:dconverge}
\left\|\frac{1}{n(\cdot,\delta_{b})}\nabla(\psi_{m}(\cdot,\delta_{a})-\psi_{m}(\cdot,\delta_{b}))\right\|_{[L^{2}(\Omega_{\mathrm{in}})]^{2}}\to0\
\text{as}\ \delta_{a}\to\delta_{b}.        
      \end{equation}      
   A direct computation yields:
   \begin{align}
     &\left\|\frac{1}{n(\cdot,\delta_{b})}\nabla(\psi_{m}(\cdot,\delta_{a})-\psi_{m}(\cdot,\delta_{b}))\right\|^{2}_{[L^{2}(\Omega_{\mathrm{in}})]^{2}}\notag\\
     =&(\frac{1}{n^{2}(\cdot,\delta_{a})}\nabla\psi_{m}(\cdot,\delta_{a}),\nabla(\psi_{m}(\cdot,\delta_{a})-\psi_{m}(\cdot,\delta_{b})))_{\Omega_{\mathrm{in}}}\notag\\
     &-(\frac{1}{n^{2}(\cdot,\delta_{b})}\nabla\psi_{m}(\cdot,\delta_{b}),\nabla(\psi_{m}(\cdot,\delta_{a})-\psi_{m}(\cdot,\delta_{b})))_{\Omega_{\mathrm{in}}}\notag\\
     &+((\frac{1}{n^{2}(\cdot,\delta_{b})}-\frac{1}{n^{2}(\cdot,\delta_{a})})\nabla\psi_{m}(\cdot,\delta_{a}),\nabla(\psi_{m}(\cdot,\delta_{a})-\psi_{m}(\cdot,\delta_{b})))_{\Omega_{\mathrm{in}}}\notag\\
     =&(\lambda_{m}(\delta_{a})\psi_{m}(\cdot,\delta_{a})-\lambda_{m}(\delta_{b})\psi_{m}(\cdot,\delta_{b}),\psi_{m}(\cdot,\delta_{a})-\psi_{m}(\cdot,\delta_{b}))_{\Omega_{\mathrm{in}}}\notag\\
      &+((\frac{1}{n^{2}(\cdot,\delta_{b})}-\frac{1}{n^{2}(\cdot,\delta_{a})})\nabla\psi_{m}(\cdot,\delta_{a}),\nabla(\psi_{m}(\cdot,\delta_{a})-\psi_{m}(\cdot,\delta_{b})))_{\Omega_{\mathrm{in}}}.\label{eq:sec3:H1est}
   \end{align}
   Here, the second equality follows from the divergence theorem and
   the equations \eqref{eq:sec2:eigen:goveq}--\eqref{eq:sec2:eigen:bc}
   satisfied by the eigenpair $(\lambda_{m},\psi_{m})$.
   Consequently, the right-hand-side of
   \eqref{eq:sec3:H1est} converges to $0$ as $\delta_{a}\to\delta_{b}$ 
   by the $L^{2}$-convergence
   $\|\psi_{m}(\cdot,\delta_{a})-\psi_{m}(\cdot,\delta_{b})\|_{L^{2}(\Omega_{\mathrm{in}})}\to0$ and   
   the assumption
   $\|n(\cdot,\delta_{a})-n(\cdot,\delta_{b})\|_{L^{\infty}(\Omega_{\mathrm{in}})}\to0$.}
  \end{proof}
\end{lemma}

In the following analysis, we always assume the eigenvalues
$\{\lambda_{m}(\delta)\}$ and eigenfunctions 
$\{\psi_{m}(\cdot,\delta)\}$ depend continuously on $\delta$, and require
\begin{equation}
  \label{eq:sec3:order}
  0=\lambda_{0}(0)<\lambda_{1}(0)\le\cdots.
\end{equation}
The following properties of $\{\lambda_{m}(\delta)\}$ and $\{\psi_{m}(\cdot,\delta)\}$
hold for each $\delta$ (cf. Theorem 4.12 in \cite{mcl00}):
  \begin{itemize}
  \item[(i).] The eigenfunctions $\{\psi_{m}\}$ can be chosen
      to be real-valued, and they form a complete
      orthonormal basis in $L^{2}(\Omega_{\mathrm{in}})$.
    \item[(ii).] $\lambda_{0}=0$,
      $\lambda_{m}>0$ for each $m\in\mathbb{Z}^{+}$ and $\lambda_{m}\to\infty$ as $m\to\infty$.
    \item[(iii).] For any $u\in{H}^{1}(\Omega_{\mathrm{in}})$, 
      \begin{align}
        &u=\sum_{m=0}^{\infty}(u,\psi_{m})_{\Omega_{\mathrm{in}}}\psi_{m}\ \text{in}\
            H^{1}(\Omega_{\mathrm{in}}),\ \text{and}\label{eq:sec3:H1expansion}\\
        &\|u\|^{2}_{H^{1}(\Omega_{\mathrm{in}})}\sim\sum_{m=0}^{\infty}(\lambda_{m}+1)|(\psi_{m},u)_{\Omega_{\mathrm{in}}}|^{2}.        \label{eq:sec3:equivalentnorm}
      \end{align}      
  \end{itemize}
Hence, we assume solutions
to \eqref{eq:sec2:goveq}--\eqref{eq:sec2:bc} attain the following
expansions in $H^{1}(\Omega_{\mathrm{in}})$:
\begin{equation}
  \label{eq:sec3:expansion:omegain}
  u=\sum_{m=0}^{\infty}d_{m}\psi_{m},\ \text{where}\ d_{m}=(u,\psi_{m})_{\Omega_{\mathrm{in}}},
\end{equation}
and adopt the $H^{1}(\Omega_{\mathrm{in}})$ norm defined in
\eqref{eq:sec3:equivalentnorm}, which is also given by
\begin{equation}
  \label{eq:sec3:equivalentnorm:integral}
  \|u\|_{H^{1}(\Omega_{\mathrm{in}})}=\sqrt{\|u\|^{2}_{L^{2}(\Omega_{\mathrm{in}})}+\|\frac{1}{n}\nabla{u}\|^{2}_{[L^{2}(\Omega_{\mathrm{in}})]^{2}}},\
  \text{for}\ u\in{H}^{1}(\Omega_{\mathrm{in}}).
\end{equation}
Moreover, \eqref{eq:sec3:equivalentnorm} implies that the sequence
$\{d_{m}\sqrt{\lambda_{m}+1}\}$ belongs to
$\ell^{2}$. Due to the conjugate symmetry of equations
\eqref{eq:sec2:goveq}--\eqref{eq:sec2:bc}, we can assume
 every BIC is real-valued, implying the sequence $\{d_{m}\}$ is also
 real-valued.
\subsection{Mode-expansion in the waveguide}
In the waveguide region $\Omega_{\mathrm{out}}:=(0,\infty)\times(-h/2,h/2)$,
we apply the method of separation of variables to the following equations:
\begin{align}
  \Delta{u}+\mu{u}=&0\ \text{in}\ \Omega_{\mathrm{out}},\label{eq:sec3:wg:govern}\\
  \partial_{\bm{\nu}}{u}=&0\ \text{on}\ \partial\Omega_{\mathrm{out}}\backslash\Gamma_{h},\label{eq:sec3:wg:bc}
\end{align}
and $u$ is supposed to be outgoing {as $x_{1}\to\infty$}. This
gives the expansions:
\begin{equation}
  \label{eq:sec3:expansion:omegaout}
  u=\sum_{j=0}^{\infty}b_{j}e^{\bi\alpha_{j}x_{1}}\phi_{j}(x_{2})\ \text{in}\ H^{1}_{\mathrm{loc}}(\Omega_{\mathrm{out}}),
\end{equation}
where $b_{j}=(u,\phi_{j})_{\Gamma_{h}}$,
\begin{equation}
  \label{eq:sec3:sepavariables}
  \phi_{j}(x_{2})=\left\{
    \begin{aligned}
      &\sqrt{\frac{1}{h}},\ &&\text{if}\ j=0;\\
      &\sqrt{\frac{2}{h}}\cos\left(\frac{j\pi}{h}x_{2}+\frac{j\pi}{2}\right),\ &&\text{if}\ j\ne0,
    \end{aligned}\right.
\end{equation}
and $\alpha_{j}=\sqrt{\mu-\left(\frac{j\pi}{h}\right)^{2}}$. We also introduce
a set of functions $\{\eta_{j}\}$ 
      \begin{equation}
        \label{eq:sec3:del:sepavariables}
        \eta_{j}:=\sqrt{\frac{2}{h}}\sin\left(\frac{j\pi}{h}x_{2}+\frac{j\pi}{2}\right)\
        \text{for}\ j\in\mathbb{Z}_{+}.
      \end{equation}
      Both $\{\phi_{j}\}$ and $\{\eta_{j}\}$ form complete orthonormal bases
      in $L^{2}(\Gamma_{h})$. Moreover, the following expansions hold:
\begin{alignat}{2}
  u|_{\Gamma_{h}}=&\sum_{j=0}^{\infty}b_{j}\phi_{j}\ &&\text{in}\ H^{1/2}(\Gamma_{h}),\label{eq:sec3:expan:Gamma:trace}\\
  \partial_{\bm{\nu}}u|_{\Gamma_{h}}=&\sum_{j=0}^{\infty}b_{j}\bi\alpha_{j}\phi_{j}\ &&\text{in}\ \widetilde{H}^{-1/2}(\Gamma_{h}).\label{eq:sec3:expan:Gamma:normal}
\end{alignat} 

\subsection{Mode-matching}
We apply the second Green identity (cf. Theorem 4.4 in \cite{mcl00})
to $u$ and $\psi_{m}$ in $\Omega_{\mathrm{in}}$ which gives
\begin{equation}
  \label{eq:sec3:secondGreen}
  (\mathcal{P}u,\psi_{m})_{\Omega_{\mathrm{in}}}-(u,\mathcal{P}\psi_{m})_{\Omega_{\mathrm{in}}}=(u,\partial_{\bm{\nu}}\psi_{m})_{\partial\Omega_{\mathrm{in}}}-(\partial_{\bm{\nu}}u,\psi_{m})_{\partial\Omega_{\mathrm{in}}}.
\end{equation}
Since $u$ and $\psi_{m}$ satisfy
\eqref{eq:sec2:goveq}--\eqref{eq:sec2:bc} and
\eqref{eq:sec2:eigen:goveq}--\eqref{eq:sec2:eigen:bc} respectively,
substituting $\partial_{\bm{v}}u$ in \eqref{eq:sec3:secondGreen} with
\eqref{eq:sec3:expan:Gamma:normal} leads to
\begin{equation}
  \label{eq:sec3:cm}
  (\lambda_{m}-\mu)d_{m}=\sum_{j=0}^{\infty}b_{j}\bi\alpha_{j}(\phi_{j},\psi_{m})_{\Gamma_{h}},\
  \text{for\ each}\ m\in\mathbb{N}.
\end{equation}
Meanwhile, matching \eqref{eq:sec3:expansion:omegain} to the
expansions \eqref{eq:sec3:expan:Gamma:trace} on $\Gamma_{h}$ and taking the
$L^{2}(\Gamma_{h})$ inner products with $\phi_{j}$, we obtain
\begin{equation}
  \label{eq:sec3:bn}
  \sum_{m=0}^{\infty}d_{m}(\psi_{m},\phi_{j})_{\Gamma_{h}}=b_{j},\
  \text{for\ each}\ j\in\mathbb{N}.
\end{equation}
Substituting \eqref{eq:sec3:bn} for $b_{j}$ in
\eqref{eq:sec3:cm}, we arrive at the following
infinite-dimensional linear system:
\begin{equation}
  \label{eq:sec3:ILS}
  \left(\begin{bmatrix}
    \lambda_{0}-\mu&&\\
               &\lambda_{1}-\mu&\\
    &&\ddots
  \end{bmatrix}-\sum_{j=0}^{\infty}\bi\alpha_{j}\mathbf{v}_{j}\mathbf{v}^{*}_{j}\right)
\begin{bmatrix}
  d_{0}\\
  d_{1}\\
  \vdots
  \end{bmatrix}=\mathbf{0},
\end{equation}
where
\begin{equation}
  \label{eq:sec3:vn}
  \mathbf{v}_{j}:=
  \begin{bmatrix}
    (\phi_{j},\psi_{0})_{\Gamma_{h}}&(\phi_{j},\psi_{1})_{\Gamma_{h}}&\cdots
  \end{bmatrix}^{T}.
\end{equation}
Hence, BICs are characterized as non-trivial solutions
$\{d_{m}\}$ to the homogeneous problem \eqref{eq:sec3:ILS}.

\section{Reduction of the system}
\label{sec:reduction}
In this section, we reduce the system \eqref{eq:sec3:ILS} to a
real-valued, finite-dimensional form by assuming $h$ is small enough
and restricting $\delta$ and $\mu$ in small intervals. 
\begin{lemma}
  \label{lem:sec4:perturbation:compatible}
  Let $M\ge3$ be an integer and suppose the eigenvalues satisfy
  \begin{equation}
  \label{eq:sec4:assumption:eigen}
  \lambda_{M-3}(0)<\lambda_{M-2}(0)\ \text{and}\ \lambda_{M-1}(0)<\lambda_{M}(0).
\end{equation}
  Then, under Condition A.1, for every $\epsilon>0$ there exists $\delta_{1}>0$
  such that for all $\delta\in[-\delta_{1},\delta_{1}]$, the following bounds hold:
  \begin{align}
    \lambda_{m}(\delta)&\le\lambda_{M-3}(0)+\epsilon,\ \forall{m\le{M-3}},    \label{eq:sec4:lam:smaller}\\
   \lambda_{m}(\delta)&\ge\lambda_{M}(0)-\epsilon,\ \ \ \ \forall{m\ge{M}}.      \label{eq:sec4:lam:greater}  
  \end{align}
  \begin{proof}
    Since only a finite number of eigenvalues exist for
    $m\le{M-3}$, \eqref{eq:sec4:lam:smaller} follows directly from the
    continuity of eigenvalues with respect to $\delta$. We focus 
    on \eqref{eq:sec4:lam:greater}. For each $m\ge{M}$,
    \begin{equation}
      \label{eq:sec4:lam:greater:0}
      \lambda_{m}(0)\ge\lambda_{M}(0)>\lambda_{M}(0)-\epsilon.
    \end{equation}
    If $\lambda_{M}(0)\le\epsilon$, then \eqref{eq:sec4:lam:greater} holds trivially.
    Assuming $\lambda_{M}(0)\ge\epsilon$, we obtain
    \begin{equation}
      \label{eq:sec4:lam:greater:0:inverse}
      \frac{1}{\lambda_{m}(0)+1}\le\frac{1}{\lambda_{M}(0)+1}<\frac{1}{\lambda_{M}(0)+1-\epsilon}.
    \end{equation}
    By selecting $\delta_{1}$ such that 
    \begin{equation}
      \label{eq:sec4:pdelta:M:inverse}
      e^{C_{0}\delta_{1}}-1\le\frac{1}{\lambda_{M}(0)+1-\epsilon}-\frac{1}{\lambda_{M}(0)+1},
    \end{equation}
    the uniform equicontinuity condition
    \eqref{eq:sec3:continuity:eigenvalue} yields the desired result.
  \end{proof}
\end{lemma}

Let $\epsilon>0$ be such that
\begin{equation}
  \label{eq:sec4:eps:settings}
  \epsilon<\frac{1}{4}\min\Big\{\lambda_{M-2}(0)-\lambda_{M-3}(0),\lambda_{M}(0)-\lambda_{M-1}(0)\Big\}.
\end{equation}
We assume $\delta$ varies within a range $[-\delta_{2},\delta_{2}]$ for some
$\delta_{2}>0$ such that
\eqref{eq:sec4:lam:smaller}--\eqref{eq:sec4:lam:greater} hold and
\begin{equation}
  \label{eq:sec4:delta:settings}  
  \lambda_{M-2}(\delta),\lambda_{M-1}(\delta)\in[\lambda_{M-3}(0)+3\epsilon,\lambda_{M}(0)-3\epsilon].
\end{equation}
Furthermore, we restrict $\mu$ to the interval
\begin{equation}
  \label{eq:sec4:k2:settings}
  \mu\in[\lambda_{M-3}(0)+2\epsilon,\lambda_{M}(0)-2\epsilon].
\end{equation}
For the subsequent analysis, we define the endpoints as
$\mu_{l}:=\lambda_{M-3}(0)+2\epsilon$ and $\mu_{r}:=\lambda_{M}(0)-2\epsilon$. We also summarize the various parameter bounds used in the
following sections as they are introduced for clarity. The intervals
are:
\begin{itemize}
\item $(0,h_{0})$: introduced in \eqref{eq:sec4:h0cond}.
\item $(0,h_{1})$: introduced in Theorem \ref{thm:sec4:L00C00:est}.
\item $(0,h_{2})$ and $[-\delta_{3},\delta_{3}]$: introduced in Theorem
  \ref{thm:sec5:f0f1}.
\item $(0,h_{3})$: introduced in Theorem
  \ref{thm:sec5:continuity:Lipcontinuity}.
\item $(0,h_{4})$: introduced in Theorem \ref{thm:sec5:BIC:existence}.  
\end{itemize}
Additionally, we require
$h_{4}\le{h_{3}}\le{h_{2}}\le{h_{1}}\le{h_{0}}$ and $\delta_{3}\le{\delta_{2}}\le\delta_{1}\le\delta_{0}$.

\subsection{Decomposition into real and imaginary components}
Letting $h\ll1$ such that $\mu<{\pi^{2}}/{h^{2}}$, only one propagating mode is
permitted in the waveguide. Hence we have
\begin{equation}
  \label{eq:sec4:ialphan}
  \alpha_{0}>0\ \text{and}\ \bi\alpha_{j}<0\ \text{for\ every}\ j\in\mathbb{Z}^{+}. 
\end{equation}
Since $\{d_{m}\}$ is real-valued, \eqref{eq:sec3:ILS} can be
decomposed by considering its real and imaginary parts separately:
\begin{align}
  &\left(\begin{bmatrix}
    \lambda_{0}-\mu&&\\
               &\lambda_{1}-\mu&\\
    &&\ddots
  \end{bmatrix}-\sum_{j=1}^{\infty}\bi\alpha_{j}\mathbf{v}_{j}\mathbf{v}^{*}_{j}\right)
\begin{bmatrix}
  d_{0}\\
  d_{1}\\
  \vdots
\end{bmatrix}=\mathbf{0},   \label{eq:sec4:ILS:real}\\
  & (\phi_{0},\sum_{m=0}^{\infty}d_{m}\psi_{m})_{\Gamma_{h}}=0.  \label{eq:sec4:ILS:imag}
\end{align}
Consequently, both \eqref{eq:sec4:ILS:real} and
\eqref{eq:sec4:ILS:imag} are real-valued.

\subsection{First reduction}
By showing a principal submatrix of \eqref{eq:sec4:ILS:real} is invertible,
we reduce \eqref{eq:sec4:ILS:real} to an $M$-dimensional linear system
in this subsection.

\begin{lemma}
  \label{lem:sec4:estimate}
  Assume $\mu<{\pi^{2}}/{h^{2}}$ and the region  
  $R_{h}:=(-h/\pi,0)\times(-h/2,h/2)$ is included in $\Omega_{\mathrm{in}}$. Then the
  following estimate holds for $u_{1},u_{2}\in{H}^{1}(\Omega_{\mathrm{in}})$:
  \begin{align}
    \left|\sum_{j=1}^{\infty}\bi\alpha_{j}(\phi_{j},u_{2})_{\Gamma_{h}}(u_{1},\phi_{j})_{\Gamma_{h}}\right|{\le}n_{1}^{2}\left\|\frac{1}{n}{\nabla}u_{1}\right\|_{[L^{2}(R_{h})]^{2}}\left\|\frac{1}{n}{\nabla}u_{2}\right\|_{[L^{2}(R_{h})]^{2}}.    \label{eq:sec4:inequality}
  \end{align}  
  \begin{proof}
    We introduce the following auxiliary functions $\chi_{j}$ for $j\in\mathbb{Z}^{+}$:
    \begin{equation}
      \label{eq:sec4:chi}
      \chi_{j}(x_{1})=\left\{
        \begin{aligned}
          &0,\ &&\text{if}\ x_{1}\le-\frac{h}{j\pi};\\
          &\frac{j\pi}{h}x_{1}+1,\ &&\text{if}\ x_{1}>-\frac{h}{j\pi}.
        \end{aligned}\right.        
    \end{equation}
    Using the divergence theorem over $R_{h}$ gives
    \begin{align}
      (u_{1},\phi_{j})_{\Gamma_{h}}=&\int_{-{h}/{2}}^{{h}/{2}}u_{1}(0,x_{2})\phi_{j}(x_{2})dx_{2}\notag\\
      =&\frac{h}{j\pi}\int_{R_{h}}\nabla\cdot({u}_{1}(x_{1},x_{2})\mathrm{curl}_{2}\left(\eta_{j}(x_{2})\chi_{j}(x_{1})\right))dx_{1}dx_{2}\notag\\
      =&\int_{R_{h}}\partial_{x_{1}}u_{1}(x_{1},x_{2})\phi_{j}(x_{2})\chi_{j}(x_{1})dx_{1}dx_{2}\notag\\
      &+\int_{R_{h}}\partial_{x_{2}}u_{1}(x_{1},x_{2})\left(-\frac{h}{j\pi}\eta_{j}(x_{2})\chi^{\prime}_{j}(x_{1})\right)dx_{1}dx_{2},\label{eq:sec4:rewrite:int}
    \end{align}
    where $\mathrm{curl}_{2}(f):=(\partial_{x_{2}}f,-\partial_{x_{1}}f)$.    
    Since both $\{\phi_{j}\}$ and $\{\eta_{j}\}$ form complete orthonormal
    bases in $L^{2}(\Gamma_{h})$, it follows that 
    $\{\sqrt{\bi\alpha_{j}}\phi_{j}\chi_{j}\}$ and
    $\{-\frac{h}{j\pi}\sqrt{\bi\alpha_{j}}\eta_{j}\chi_{j}^{\prime}\}$ are orthogonal
    sequences in $L^{2}(R_{h})$. Moreover, their norms are uniformly bounded with respect to
    $j$ that
    \begin{align}
      \left\|\sqrt{\bi\alpha_{j}}\phi_{j}\chi_{j}\right\|_{L^{2}(R_{h})}^{2}=&\frac{1}{3}\sqrt{1-\frac{h^{2}}{\pi^{2}j^{2}}\mu}<1,\label{eq:sec4:norm1}\\
      \left\|-\frac{h}{j\pi}\sqrt{\bi\alpha_{j}}\eta_{j}\chi_{j}^{\prime}\right\|_{L^{2}(R_{h})}^{2}=&\sqrt{1-\frac{h^{2}}{\pi^{2}j^{2}}\mu}<1.\label{eq:sec4:norm2}      
    \end{align}
Hence, \eqref{eq:sec4:inequality} comes from the Cauchy-Schwartz
inequality and Bessel's inequality.
  \end{proof}
\end{lemma}

Let $h_{0}>0$ be chosen so that 
\begin{equation}
  \label{eq:sec4:h0cond}
  \mu_{r}<\lambda_{M}(0)-\epsilon<\frac{\pi^{2}}{h_{0}^{2}}\ \text{and}\ R_{h_{0}}\subset\Omega_{B}.
\end{equation}
In the following analysis, we always assume
$h\in(0,h_{0})$. To reduce the system
\eqref{eq:sec4:ILS:real}--\eqref{eq:sec4:ILS:imag} into a finite-dimensional
form,
we adopt the following notation:
\begin{align}
  \mathbf{A}:=&\sum_{j=1}^{\infty}\bi\alpha_{j}\mathbf{v}_{0,M-1,j}\mathbf{v}_{0,M-1,j}^{*},\label{eq:sec4:matrix:A}\\
  \mathbf{B}:=&\mathbf{D}\left(\sum_{j=1}^{\infty}\bi\alpha_{j}\mathbf{v}_{M,\infty,j}\mathbf{v}_{M,\infty,j}^{*}\right)\mathbf{D},\label{eq:sec4:matrix:B}\\
    \mathbf{V}:=&\mathbf{D}\sum_{j=1}^{\infty}\bi\alpha_{j}
                      \mathbf{v}_{M,\infty,j}\mathbf{v}_{0,M-1,j}^{*},\label{eq:sec4:matrix:V}
\end{align}
where
\begin{equation*}
  \mathbf{v}_{M_{1},M_{2},j}:=
    \begin{bmatrix}
      (\phi_{j},\psi_{M_{1}})_{\Gamma_{h}}\\\vdots\\(\phi_{j},\psi_{M_{2}})_{\Gamma_{h}}
    \end{bmatrix},\
    \mathbf{D}:=
  \begin{bmatrix}
    \frac{1}{\sqrt{\lambda_{M}-\mu}}&&\\
                                &\frac{1}{\sqrt{\lambda_{M+1}-\mu}}&\\
                                &&\ddots
  \end{bmatrix},
\end{equation*}
and $\mathbf{v}_{M_{1},\infty,j}$ denotes the infinite-dimensional vector $\mathbf{v}_{M_{1},M_{2},j}$ as $M_{2}\to\infty$.
\begin{theorem}
  \label{thm:sec4:finite:sys}  
  Under Condition A.1 and the assumption $h\in(0,h_{0})$, the operator
  $\mathbf{B}$ is  
  uniformly bounded in $\ell^{2}$ for $\delta\in[-\delta_{2},\delta_{2}]$ and
  $\mu\in[\mu_{l},\mu_{r}]$. Furthermore, we can obtain  
  \begin{equation}
    \label{eq:sec4:I-A1:est}
   \|(\mathbf{I}-\mathbf{B})^{-1}\|_{\ell^{2}\to\ell^{2}}\le1,
 \end{equation}
 where $\mathbf{I}$ denotes the identity operator.
 Additionally, for each $m=0,\ldots,M-1$, the vector
  $\sum_{j=1}^{\infty}\bi\alpha_{j}(\phi_{j},\psi_{m})_{\Gamma_{h}}\mathbf{v}_{M,\infty,j}^{*}\mathbf{D}$
  belongs to $\ell^{2}$ within these parameter ranges, with its norm
  bounded by
  \begin{equation}
    \label{eq:sec4:V:est}
   \left\|\sum_{j=1}^{\infty}\bi\alpha_{j}(\phi_{j},\psi_{m})_{\Gamma_{h}}\mathbf{v}_{M,\infty,j}^{*}\mathbf{D}\right\|_{\ell^{2}}\le{n_{1}^{2}}\sqrt{1+\frac{{\lambda_{M}(0)+1}}{\epsilon}}\left\|\frac{1}{n}\nabla\psi_{m}\right\|_{[L^{2}(R_{h})]^{2}}.
 \end{equation}
 \begin{proof}
   Let $\{d_{1,m}\}_{m=M}^{\infty}$ and
   $\{d_{2,m}\}_{m=M}^{\infty}$ denote sequences in $\ell^{2}$. A direct
   computation gives
   \begin{align}
     \label{eq:sec4:A1:norm}
     &(\mathbf{B}\{d_{1,m}\},\{d_{2,m}\})_{\ell^{2}}\notag\\
     =&\sum_{j=1}^{\infty}\bi\alpha_{j}(\phi_{j},\sum_{m=M}^{\infty}\frac{d_{1,m}}{\sqrt{\lambda_{m}-\mu}}\psi_{m})_{\Gamma_{h}}(\sum_{m=M}^{\infty}\frac{d_{2,m}}{\sqrt{\lambda_{m}-\mu}}\psi_{m},\phi_{j})_{\Gamma_{h}}\notag\\
     \le&n_{1}^{2}\left\|\sum_{m=M}^{\infty}\frac{d_{1,m}\psi_{m}}{\sqrt{\lambda_{m}-\mu}}\right\|_{H^{1}(R_{h})}\left\|\sum_{m=M}^{\infty}\frac{d_{2,m}\psi_{m}}{\sqrt{\lambda_{m}-\mu}}\right\|_{H^{1}(R_{h})}\notag\\
     \le&n_{1}^{2}\sqrt{\sum_{m=M}^{\infty}\left(1+\frac{\mu+1}{\lambda_{m}-\mu}\right)|d_{1,m}|^{2}}\sqrt{\sum_{m=M}^{\infty}\left(1+\frac{\mu+1}{\lambda_{m}-\mu}\right)|d_{2,m}|^{2}}\notag\\
     \le&n_{1}^{2}\left(1+\frac{\lambda_{M}(0)+1}{\epsilon}\right)\|\{d_{1,m}\}\|_{\ell^{2}}\|\{d_{2,m}\}\|_{\ell^{2}},
   \end{align}
where the first inequality follows from Lemma \ref{lem:sec4:estimate}
and the second inequality follows from our definition of
$H^{1}(\Omega_{\mathrm{in}})$ norm in \eqref{eq:sec3:equivalentnorm}. Since
$-\mathbf{B}$ is positive, 
it follows that
\begin{equation}
  \label{eq:sec4:positive}
  ((\mathbf{I}-\mathbf{B})\mathbf{v},\mathbf{v})_{\ell^{2}}\ge\|\mathbf{v}\|_{\ell^{2}}^{2}\
  \text{for\ every}\ \mathbf{v}\in\ell^{2}.
\end{equation}
Then the invertibility of
$\mathbf{I}-\mathbf{B}$ comes from the lemma of Lax and Milgram 
(cf. Lemma 2.32 in \cite{mcl00}).
Letting
$\widetilde{\mathbf{v}}:=(\mathbf{I}-\mathbf{B})\mathbf{v}$, the
equation \eqref{eq:sec4:positive} implies
\begin{equation}
  \label{eq:sec4:positive:2}
  \|(\mathbf{I}-\mathbf{B})^{-1}\|_{\ell^{2}\to\ell^{2}}\|\widetilde{\mathbf{v}}\|^{2}_{\ell^{2}}\ge(\widetilde{\mathbf{v}},(\mathbf{I}-\mathbf{B})^{-1}\widetilde{\mathbf{v}})_{\ell^{2}}\ge\|(\mathbf{I}-\mathbf{B})^{-1}\widetilde{\mathbf{v}}\|_{\ell^{2}}^{2},
\end{equation}
which leads to \eqref{eq:sec4:I-A1:est}.
Finally, the estimate
\eqref{eq:sec4:V:est} can be shown following the same procedures presented
in \eqref{eq:sec4:A1:norm}.
 \end{proof}
\end{theorem}

Consider the subsystem in \eqref{eq:sec4:ILS:real} satisfied by
$[d_{M},d_{M+1},\ldots]^{T}$:
\begin{equation}
  \label{eq:sec4:cM}
  (\mathbf{I}-\mathbf{B})\mathbf{D}^{-1}\begin{bmatrix}
    d_{M}\\
    d_{M+1}\\
    \vdots
  \end{bmatrix}=\mathbf{V}
  \begin{bmatrix}
    d_{0}\\
    \vdots\\
    d_{M-1}
  \end{bmatrix}.
\end{equation}
Theorem \ref{thm:sec4:finite:sys} implies $[d_{M},d_{M+1},\ldots]^{T}$ can
be represented by $[d_{0},\cdots,d_{M-1}]^{T}$, which leads to 
a finite dimensions linear system:
  \begin{align}
    &\left(\begin{bmatrix}
      \lambda_{0}-\mu&&\\
                 &\ddots&\\
      &&\lambda_{M-1}-\mu
    \end{bmatrix}-\mathbf{A}-\mathbf{V}^{*}(\mathbf{I}-\mathbf{B})^{-1}\mathbf{V}\right)
  \begin{bmatrix}
    d_{0}\\
    \vdots\\
    d_{M-1}
  \end{bmatrix}=\mathbf{0};    \label{eq:sec4:FLS:real}\\
  &\left(\mathbf{v}_{0,M-1,0}^{*}+\mathbf{v}_{M,\infty,0}^{*}\mathbf{D}(\mathbf{I}-\mathbf{B})^{-1}\mathbf{V}\right)  \begin{bmatrix}
    d_{0}\\
    \vdots\\
    d_{M-1}
  \end{bmatrix}=0.  \label{eq:sec4:FLS:imag}
\end{align}

\subsection{Further reduction}
We further reduce the system
\eqref{eq:sec4:FLS:real}–\eqref{eq:sec4:FLS:imag} to a $3\times2$ linear
system for the vector $[d_{M-2},d_{M-1}]^{T}$. By evaluating the
determinants of two submatrices, we derive two governing
equations. The existence of BICs corresponds to determining pairs
$(\delta,\mu)$ for which both equations are satisfied. To
formalize this, we introduce the following notation:
\begin{align}
  \mathbf{C}_{0,0}:=&\sum_{j=1}^{\infty}\bi\alpha_{j}\mathbf{v}_{0,M-3,j}\mathbf{v}_{0,M-3,j}^{*}\notag\\
                    &+\sum_{j=1}^{\infty}\bi\alpha_{j}\mathbf{v}_{0,M-3,j}\mathbf{v}_{M,\infty,j}^{*}\mathbf{D}(\mathbf{I}-\mathbf{B})^{-1}\mathbf{D}\sum_{j=1}^{\infty}\bi\alpha_{j}\mathbf{v}_{M,\infty,j}\mathbf{v}^{*}_{0,M-3,j},\label{eq:sec4:mat:C00}\\
  \mathbf{C}_{0,1}:=&\sum_{j=1}^{\infty}\bi\alpha_{j}\mathbf{v}_{0,M-3,j}\mathbf{v}^{*}_{M-2,M-1,j}\notag\\
                    &+\sum_{j=1}^{\infty}\bi\alpha_{j}\mathbf{v}_{0,M-3,j}\mathbf{v}_{M,\infty,j}^{*}\mathbf{D}(\mathbf{I}-\mathbf{B})^{-1}\mathbf{D}\sum_{j=1}^{\infty}\bi\alpha_{j}\mathbf{v}_{M,\infty,j}\mathbf{v}^{*}_{M-2,M-1,j},\label{eq:sec4:mat:C01}\\
  \mathbf{C}_{1,0}:=&\mathbf{C}_{0,1}^{*},\label{eq:sec4:mat:C10}\\  
  \mathbf{C}_{1,1}:=&\sum_{j=1}^{\infty}\bi\alpha_{j}\mathbf{v}_{M-2,M-1,j}\mathbf{v}_{M-2,M-1,j}^{*}\notag\\
                    &+\sum_{j=1}^{\infty}\bi\alpha_{j}\mathbf{v}_{M-2,M-1,j}\mathbf{v}_{M,\infty,j}^{*}\mathbf{D}(\mathbf{I}-\mathbf{B})^{-1}\mathbf{D}\sum_{j=1}^{\infty}\bi\alpha_{j}\mathbf{v}_{M,\infty,j}\mathbf{v}^{*}_{M-2,M-1,j},\label{eq:sec4:mat:C11}\\
  \mathbf{L}_{0,0}:=&\begin{bmatrix}
                       \lambda_{0}-\mu&&\\
                                  &\ddots&\\
                       &&\lambda_{M-3}-\mu
                      \end{bmatrix},\ \text{and}\ \mathbf{L}_{1,1}:=
                                                          \begin{bmatrix}
                                                            \lambda_{M-2}-\mu&\\
                                                            &\lambda_{M-1}-\mu
                                                          \end{bmatrix}.\label{eq:sec4:L00:L11}
\end{align}
Hence the linear system \eqref{eq:sec4:FLS:real} takes the form:
  \begin{equation}
    \label{eq:sec4:reduceLS:2}
    \left(\begin{bmatrix}
      \mathbf{L}_{0,0}&\\
      &\mathbf{L}_{1,1}
          \end{bmatrix}-
     \begin{bmatrix}
      \mathbf{C}_{0,0}&\mathbf{C}_{0,1}\\
      \mathbf{C}_{1,0}&\mathbf{C}_{1,1}
    \end{bmatrix}\right)
  \begin{bmatrix}
    d_{0}\\
    \vdots\\
    d_{M-1}
  \end{bmatrix}=\mathbf{0}.
\end{equation}

\begin{lemma}
  \label{lem:sec4:C00:est}
  Under Condition A.1 and the assumption $h\in(0,h_{0})$, each entry in
  the matrices $\mathbf{C}_{0,0},\ \mathbf{C}_{0,1},\
  \mathbf{C}_{1,0}$ and $\mathbf{C}_{1,1}$ converges to $0$ uniformly
  as $h\to0$ for $\delta\in[-\delta_{2},\delta_{2}]$ and $\mu\in[\mu_{l},\mu_{r}]$.  
  \begin{proof}
    We demonstrate this result for each entry in
    $\mathbf{C}_{0,0}$. Let $c_{m_{1},m_{2}}$ denote the entry in the
    $m_{1}$-th row and    
    $m_{2}$-th column of $\mathbf{C}_{0,0}$. Then $c_{m_{1},m_{2}}$
    can be expressed as:
    \begin{align}
      c_{m_{1},m_{2}}=&\sum_{j=1}^{\infty}\bi\alpha_{j}(\phi_{j},\psi_{m_{1}})_{\Gamma_{h}}(\psi_{m_{2}},\phi_{j})_{\Gamma_{h}}\notag\\
                      &+\sum_{j=1}^{\infty}\bi\alpha_{j}(\phi_{j},\psi_{m_{1}})_{\Gamma_{h}}\mathbf{v}_{M,\infty,j}^{*}\mathbf{D}(\mathbf{I}-\mathbf{B})^{-1}\mathbf{D}\sum_{j=1}^{\infty}\bi\alpha_{j}\mathbf{v}_{M,\infty,j}(\psi_{m_{2}},\phi_{j})_{\Gamma_{h}}.\label{eq:sec4:cm1m2:exp}
    \end{align}
    Using Lemma \ref{lem:sec4:estimate} and Theorem
    \ref{thm:sec4:finite:sys}, we can obtain the bound:
    \begin{align}
    |c_{m_{1},m_{2}}|\le&n_{1}^{2}\left\|\frac{1}{n}{\nabla}\psi_{m_{1}}\right\|_{[L^{2}(R_{h})]^{2}}\left\|\frac{1}{n}{\nabla}\psi_{m_{2}}\right\|_{[L^{2}(R_{h})]^{2}}\notag\\
      &+{n_{1}^{4}}\left(1+\frac{{\lambda_{M}(0)+1}}{\epsilon}\right)\left\|\frac{1}{n}\nabla\psi_{m_{1}}\right\|_{[L^{2}(R_{h})]^{2}}\left\|\frac{1}{n}\nabla\psi_{m_{2}}\right\|_{[L^{2}(R_{h})]^{2}}.\label{eq:sec4:cm1m2:est}
    \end{align}
    Thus, it's sufficient to show the uniform convergence:
    \begin{equation*}
      \label{eq:sec4:psim1}
      \left\|\frac{1}{n}\nabla\psi_{m}(\cdot,\delta)\right\|_{[L^{2}(R_{h})]^{2}}\to0\ \text{as}\
      h\to0,\ \text{for}\ m=m_{1},m_{2}.
    \end{equation*}
    For fixed $h$, consider the estimate for $\delta_{a},\delta_{b}\in[-\delta_{2},\delta_{2}]$:
    \begin{align}
      &\left|\left\|\frac{1}{n}\nabla\psi_{m}(\cdot,\delta_{a})\right\|_{[L^{2}(R_{h})]^{2}}-\left\|\frac{1}{n}\nabla\psi_{m}(\cdot,\delta_{b})\right\|_{[L^{2}(R_{h})]^{2}}\right|\notag\\
      \le&\left\|\frac{1}{n}\left(\nabla\psi_{m}(\cdot,\delta_{a})-\nabla\psi_{m}(\cdot,\delta_{b})\right)\right\|_{[L^{2}(R_{h})]^{2}}\notag\\
      \le&\|\psi_{m}(\cdot,\delta_{a})-\psi_{m}(\cdot,\delta_{b})\|_{H^{1}(\Omega_{\mathrm{in}})}.\label{eq:sec4:psim1:est}      
    \end{align}
    By \eqref{eq:sec3:H1converge}, the right hand side converges to
    $0$ as    
    $\delta_{a}\to\delta_{b}$, establishing the uniform equicontinuity of 
    $\|\frac{1}{n}\nabla\psi_{m}(\cdot,\delta)\|_{[L^{2}(R_{h})]^{2}}$ for all
    $h\in(0,h_{0})$. The Arzel$\grave{\mathrm{a}}$-Ascoli    
    theorem then guarantees the uniform convergence of
    $\|\frac{1}{n}\nabla\psi_{m}(\cdot,\delta)\|_{[L^{2}(R_{h})]^{2}}$ to $0$ as $h\to0$.
  \end{proof}
\end{lemma}

\begin{theorem}
  \label{thm:sec4:L00C00:est}
  Under Condition A.1, there exists $h_{1}>0$ such
  that the matrix $\mathbf{L}_{0,0}-\mathbf{C}_{0,0}$ is invertible
  for all  
  $h\in(0,h_{1})$, $\delta\in[-\delta_{2},\delta_{2}]$ and $\mu\in[\mu_{l},\mu_{r}]$. Moreover, there
  exists a constant $C_{1}>1/\epsilon$ such that  
  \begin{equation}
    \label{eq:sec4:L00C00:norm}
   \|(\mathbf{L}_{0,0}-\mathbf{C}_{0,0})^{-1}\|_{2}\le{C_{1}}.
  \end{equation}
  \begin{proof}
    Under the assumed ranges of $\delta$ and $\mu$,
    the matrix $\mathbf{L}_{0,0}$ is negative definite with
    eigenvalues less than    
    $-\epsilon$, ensuring its invertibility. By Lemma
    \ref{lem:sec4:C00:est}, the operator norm $\|\mathbf{C}_{0,0}\|_{2}\to0$
    as $h\to0$. Choose $h_{1}<h_{0}$ such that
    $\|\mathbf{C}_{0,0}\|_{2}\le\epsilon-1/C_{1}$ for all $h<h_{1}$. Then, by the
    Neumann series expansion, we obtain:
    \begin{align*}
      \|(\mathbf{L}_{0,0}-\mathbf{C}_{0,0})^{-1}\|_{2}\le&\|\mathbf{L}_{0,0}^{-1}\|_{2}\|(\mathbf{I}-\mathbf{L}_{0,0}^{-1}\mathbf{C}_{0,0})^{-1}\|_{2}\\
      \le&\|\mathbf{L}_{0,0}^{-1}\|_{2}\sum_{j=0}^{\infty}\|\mathbf{L}_{0,0}^{-1}\mathbf{C}_{0,0}\|_{2}^{n}\le\frac{1}{\epsilon-\|\mathbf{C}_{0,0}\|}\le{C_{1}}.
    \end{align*}
  \end{proof}
\end{theorem}

Utilizing Theorem \ref{thm:sec4:L00C00:est}, the coefficients
$[d_{0},\ldots,d_{M-3}]^{T}$ can be determined from
\begin{equation}
  \label{eq:sec4:C0CM-3}
  \begin{bmatrix}
    d_{0}\\
    \vdots\\
    d_{M-3}
  \end{bmatrix}=(\mathbf{L}_{0,0}-\mathbf{C}_{0,0})^{-1}\mathbf{C}_{0,1}
  \begin{bmatrix}
    d_{M-2}\\
    d_{M-1}
  \end{bmatrix}.
\end{equation}
Hence the system
\eqref{eq:sec4:FLS:real}--\eqref{eq:sec4:FLS:imag} can be further reduced to 
\begin{align}  
  &\left(\mathbf{L}_{1,1}-\mathbf{C}_{1,1}-\mathbf{C}_{1,0}(\mathbf{L}_{0,0}-\mathbf{C}_{0,0})^{-1}\mathbf{C}_{0,1}\right)
  \begin{bmatrix}
    d_{M-2}\\
    d_{M-1}
  \end{bmatrix}=\mathbf{0};\label{eq:sec4:RFLS:real}\\
  &\Big(\mathbf{v}_{M-2,M-1,0}^{*}+\mathbf{v}_{0,M-3,0}^{*}(\mathbf{L}_{0,0}-\mathbf{C}_{0,0})^{-1}\mathbf{C}_{0,1}\notag\\
&\qquad\qquad  +\mathbf{v}_{M,\infty,0}^{*}\mathbf{D}(\mathbf{I}-\mathbf{B})^{-1}\mathbf{V}
     \begin{bmatrix}
       (\mathbf{L}_{0,0}-\mathbf{C}_{0,0})^{-1}\mathbf{C}_{0,1}\\
       \mathbf{I}_{2}
     \end{bmatrix}\Big)
     \begin{bmatrix}
       d_{M-2}\\
       d_{M-1}
     \end{bmatrix}=0,\label{eq:sec4:RFLS:imag}  
\end{align}
where $\mathbf{I}_{2}$ denotes the identity operator in
$\mathbb{R}^{2}$. Consequently, BICs serve as non-trivial
solutions to the system \eqref{eq:sec4:RFLS:real}--\eqref{eq:sec4:RFLS:imag}, which can be summarized as follows:
\begin{equation}
  \label{eq:sec4:simLS}
  \begin{bmatrix}
    \lambda_{M-2}-\mu+a_{0,0}&a_{0,1}\\
    a_{1,0}&\lambda_{M-1}-\mu+a_{1,1}\\
    a_{2,0}&a_{2,1}
  \end{bmatrix}
  \begin{bmatrix}
    d_{M-2}\\
    d_{M-1}
  \end{bmatrix}=\mathbf{0}.
\end{equation}
Here $a_{0,0}$, $a_{0,1}$, $a_{1,0}$ and $a_{1,1}$ denote the entries of
$-\mathbf{C}_{1,1}-\mathbf{C}_{1,0}(\mathbf{L}_{0,0}-\mathbf{C}_{0,0})^{-1}\mathbf{C}_{0,1}$,
while $a_{2,0}$ and $a_{2,1}$ 
correspond to the coefficients in \eqref{eq:sec4:RFLS:imag}.
Thus, the existence of BICs is equivalent to finding solutions to the following equations:
\begin{align}
  \lambda_{M-2}-\mu+a_{0,0}-a_{0,1}\frac{a_{2,0}}{a_{2,1}}&=0, \label{eq:sec4:exBIC:cond1}\\
  \lambda_{M-1}-\mu+a_{1,1}-a_{1,0}\frac{a_{2,1}}{a_{2,0}}&=0, \label{eq:sec4:exBIC:cond2}
\end{align}
which can be derived by computing the determinants of two submatrices from \eqref{eq:sec4:simLS}.

\section{Existence of BICs under analytic medium perturbations}
\label{sec:existenceofFWBICs}
In this section, 
we introduce the notation of functions $f_{0}(\delta,\mu,h)$ and
$f_{1}(\delta,\mu,h)$, defined as:
\begin{align}
    f_{0}(\delta,\mu,h)=a_{0,0}-a_{0,1}\frac{a_{2,0}}{a_{2,1}},\label{eq:sec5:f0}\\
    f_{1}(\delta,\mu,h)=a_{1,1}-a_{1,0}\frac{a_{2,1}}{a_{2,0}}.\label{eq:sec5:f1}
\end{align}  
Building on Conditions A.1 and A.2 from Section
\ref{sec2:MainContribution} and $h\ll1$, we establish the following
properties of $f_{0}$ and $f_{1}$ for $\delta$ and $\mu$ in small intervals:
\begin{itemize}
    \item uniform convergence: $f_{0},f_{1}\to0$ as $h\to0$ uniformly;
    \item continuity in $\delta$: $f_{0}$ and $f_{1}$ are continuous with
      respect to $\delta$;
    \item Lipschitz continuity in $\mu$: both $f_{0}$ and $f_{1}$ are
      Lipschitz continuous in $\mu$ with a Lipschitz constant less than 1.
\end{itemize}
Consequently, each of the equations
\eqref{eq:sec4:exBIC:cond1}--\eqref{eq:sec4:exBIC:cond2} defines an
implicit function $\mu(\delta)$. BICs emerge as intersections of the
graphs of these implicit functions, and their existence is guaranteed
under Condition A.3.

\subsection{Uniform convergence of $f_{0}$ and $f_{1}$ to $0$ as
  $h\to0$}
We show $f_{0}(\delta,\mu,h)$ and $f_{1}(\delta,\mu,h)$ are well-defined and
approach $0$ as $h\to0$ uniformly over small intervals of $\delta$ and $\mu$.

\begin{lemma}
  \label{lem:sec5:H3converge}
  Let $\Omega_{b}\subset{\Omega_{\mathrm{in}}}$ be a region with a $C^{1}$ boundary satisfying
  $\overline{\Omega_{b}}\subset\Omega_{B}\cup\{x_{1}=0\}$ and $\Gamma_{h}\subset\partial{\Omega_{b}}$.
  Under Condition A.1, the mapping
  $\delta\to\psi_{m}(\cdot,\delta)|_{\Omega_{b}}$ is continuous from  
  $[-\delta_{0},\delta_{0}]$ to $H^{3}(\Omega_{b})$ for each $m\in\mathbb{N}$, i.e.,
  \begin{equation}
    \label{eq:sec5:H3converge}
    \|\psi_{m}(\cdot,\delta_{a})-\psi_{m}(\cdot,\delta_{b})\|_{H^{3}(\Omega_{b})}\to0\
    \text{as}\ \delta_{a}\to\delta_{b},\ \text{for}\ \delta_{a},\delta_{b}\in[-\delta_{0},\delta_{0}].
  \end{equation}  
  \begin{proof} 
    For each $m\in\mathbb{N}$, observe that $\lambda_{m}$ and $\psi_{m}$ satisfy the eigenvalue
    problem \eqref{eq:sec2:eigen:goveq}--\eqref{eq:sec2:eigen:bc}. From
    this, we obtain:
    \begin{equation}
      -\Delta\psi_{m}(\mathbf{x},\delta)=\lambda_{m}(\delta)\psi_{m}(\mathbf{x},\delta),\ \text{for}\ \mathbf{x}\in\Omega_{B}.\label{eq:sec5:laplacepsi}
    \end{equation}
    The mapping $\delta\to-\Delta\psi_{m}(\cdot,\delta)|_{\Omega_{B}}$ is therefore continuous from
    $[-\delta_{0},\delta_{0}]$ to $H^{1}(\Omega_{B})$ by using
    Lemma \ref{lem:sec3:eigen:continuity}. Since
    $\partial_{\bm{\nu}}\psi_{m}(\cdot,\delta)=0$ on $\partial\Omega_{\mathrm{in}}$, the regularity
    theory ensures that 
    $\psi_{m}(\cdot,\delta)\in{H^{3}(\Omega_{b})}$ (cf. Theorem 4.18 in
    \cite{mcl00}). Furthermore, for 
    $\delta_{a},\delta_{b}\in[-\delta_{0},\delta_{0}]$, the estimate 
\begin{align}
  \|\psi_{m}(\cdot,\delta_{a})-\psi_{m}(\cdot,\delta_{b})\|_{H^{3}(\Omega_{b})}{\le}&C\|\psi_{m}(\cdot,\delta_{a})-\psi_{m}(\cdot,\delta_{b})\|_{H^{1}(\Omega_{B})}\notag\\         
  &+C\|\Delta\psi_{m}(\cdot,\delta_{a})-\Delta\psi_{m}(\cdot,\delta_{b})\|_{H^{1}(\Omega_{B})}.
   \label{eq:sec5:regularity}
\end{align}
holds for a constant $C>0$.
Finally, \eqref{eq:sec5:H3converge} follows by combining
\eqref{eq:sec3:H1converge} with the continuity of
$\delta\to-\Delta\psi_{m}(\cdot,\delta)|_{\Omega_{B}}$ in $H^{1}(\Omega_{B})$.
  \end{proof}
\end{lemma}

\begin{theorem}
  \label{thm:sec5:f0f1}
  Under Conditions A.1 and A.2, there exist constants $h_{2},\delta_{3}>0$
  such that for all $h\in(0,h_{2})$,
  $\delta\in[-\delta_{3},\delta_{3}]$ and $\mu\in[\mu_{l},\mu_{r}]$,
  the functions $f_{0}$ and $f_{1}$ are well-defined.
  Moreover, these functions vanish uniformly in the limit:
\begin{equation}
  \label{eq:sec5:f0f1:to0}
  \lim_{h\to0}f_{0}(\cdot,\cdot,h)=0,\ \text{and}\ \lim_{h\to0}f_{1}(\cdot,\cdot,h)=0.
\end{equation}
\begin{proof}
By Lemma \ref{lem:sec4:C00:est} and Theorem
\ref{thm:sec4:L00C00:est}, the terms $a_{0,0}$, $a_{0,1}$, $a_{1,0}$ and
$a_{1,1}$ are all well-defined and converge uniformly to $0$ as $h\to0$.
The proof therefore reduces to showing that the ratios
${a_{2,0}}/{a_{2,1}}$ and ${a_{2,1}}/{a_{2,0}}$ are uniformly bounded,
from which \eqref{eq:sec5:f0f1:to0} follows.
From \eqref{eq:sec4:RFLS:imag}, we express $a_{2,0}$ as:
  \begin{align}
    a_{2,0}=&(\psi_{M-2},\phi_{0})_{\Gamma_{h}}+\mathbf{v}_{0,M-3,0}^{*}(\mathbf{L}_{0,0}-\mathbf{C}_{0,0})^{-1}\mathbf{q}\notag\\
            &+\mathbf{v}_{M,\infty,0}^{*}\mathbf{D}(\mathbf{I}-\mathbf{B})^{-1}\mathbf{D}\sum_{j=1}^{\infty}\bi\alpha_{j}\mathbf{v}_{M,\infty,j}\mathbf{v}_{0,M-3,j}^{*}(\mathbf{L}_{0,0}-\mathbf{C}_{0,0})^{-1}\mathbf{q}\notag\\
    &+\mathbf{v}_{M,\infty,0}^{*}\mathbf{D}(\mathbf{I}-\mathbf{B})^{-1}\mathbf{D}\sum_{j=1}^{\infty}\bi\alpha_{j}\mathbf{v}_{M,\infty,j}(\psi_{M-2},\phi_{j})_{\Gamma_{h}},
    \label{eq:sec5:a20}    
  \end{align}
  with the vector $\mathbf{q}$ given by:
  \begin{align}
    \mathbf{q}:=&\sum_{j=1}^{\infty}\bi\alpha_{j}(\phi_{j},\psi_{M-2})_{\Gamma_{h}}\mathbf{v}_{0,M-3,j}\notag\\
                &+\sum_{j=1}^{\infty}\bi\alpha_{j}\mathbf{v}_{0,M-3,j}\mathbf{v}^{*}_{M,\infty,j}\mathbf{D}(\mathbf{I}-\mathbf{B})^{-1}\mathbf{D}\sum_{j=1}^{\infty}\bi\alpha_{j}\mathbf{v}_{M,\infty,j}(\Psi_{M-2},\phi_{j})_{\Gamma_{h}}.
  \end{align}
An application of Lemma \ref{lem:sec4:estimate} and
Theorem \ref{thm:sec4:finite:sys} yields the estimate:
\begin{align}
  \|\mathbf{q}\|_{2}\le&{n_{1}^{2}\left\|\frac{1}{n}{\nabla}\psi_{M-2}\right\|_{[L^{2}(R_{h})]^{2}}\left(\sum_{m=0}^{M-3}\left\|\frac{1}{n}{\nabla}\psi_{m}\right\|^{2}_{[L^{2}(R_{h})]^{2}}\right)^{1/2}}\notag\\
 &+{n_{1}^{4}}\left(1+\frac{{\lambda_{M}(0)+1}}{\epsilon}\right)\left\|\frac{1}{n}\nabla\psi_{M-2}\right\|_{[L^{2}(R_{h})]^{2}}\left(\sum_{m=0}^{M-3}\left\|\frac{1}{n}{\nabla}\psi_{m}\right\|^{2}_{[L^{2}(R_{h})]^{2}}\right)^{1/2}.
\end{align}
Using this and Theorem \ref{thm:sec4:L00C00:est}, we obtain
  \begin{align}
    &|a_{2,0}-(\psi_{M-2},\phi_{0})_{\Gamma_{h}}|\notag\\
    \le&C_{1}\|\mathbf{v}_{0,M-3,0}^{*}\|_{2}\|\mathbf{q}\|_{2}\notag\\
    &+C_{1}n_{1}^{2}\sqrt{1+\frac{{\lambda_{M}(0)+1}}{\epsilon}}\left(\sum_{m=0}^{M-3}\left\|\frac{1}{n}{\nabla}\psi_{m}\right\|^{2}_{[L^{2}(R_{h})]^{2}}\right)^{1/2}\|\mathbf{v}_{M,\infty,0}^{*}\mathbf{D}\|_{\ell^{2}}\|\mathbf{q}\|_{2}\notag\\
    &+n_{1}^{2}\sqrt{1+\frac{{\lambda_{M}(0)+1}}{\epsilon}}\left\|\frac{1}{n}\nabla\psi_{M-2}\right\|_{[L^{2}(R_{h})]^{2}}\|\mathbf{v}_{M,\infty,0}^{*}\mathbf{D}\|_{\ell^{2}}.
    \label{eq:sec5:a20:est}    
  \end{align}
  Consider the
  sequence $\{d_{0,m}\}_{m=M}^{\infty}\in\ell^{2}$. The inner product
  with $\mathbf{v}^{*}_{M,\infty,0}\mathbf{D}$ satisfies:
  \begin{align}
    \left|(\mathbf{v}^{*}_{M,\infty,0}\mathbf{D},\{d_{0,m}\})_{\ell^{2}}\right|=&\left|(\phi_{0},\sum_{m=M}^{\infty}\frac{d_{0,m}}{\sqrt{\lambda_{m}-\mu}}\psi_{m})_{\Gamma_{h}}\right|\notag\\
    \le&\left\|\sum_{m=M}^{\infty}\frac{d_{0,m}}{\sqrt{\lambda_{m}-\mu}}\psi_{m}\right\|_{L^{2}(\Gamma_{h})}
    {\le}C\left\|\sum_{m=M}^{\infty}\frac{d_{0,m}}{\sqrt{\lambda_{m}-\mu}}\psi_{m}\right\|_{H^{1}(\Omega_{\mathrm{in}})}\notag\\
    \le&Cn_{1}^{2}\sqrt{1+\frac{{\lambda_{M}(0)+1}}{\epsilon}}\|\{d_{0,m}\}\|_{\ell^{2}},
  \end{align}
where $C$ is a constant independent of $h$. Hence, 
$\mathbf{v}_{0,M-3,0}^{*}$ and $\mathbf{v}_{M,\infty,0}^{*}\mathbf{D}$
are uniformly bounded. For small $h$, $R_{h}$ lies within
$\Omega_{b}$ as defined in Lemma \ref{lem:sec5:H3converge}.
By the general
Sobolev inequalities (cf. Section 5 in \cite{evans98}), the
convergence in \eqref{eq:sec5:H3converge} implies that the mapping
$\delta\to\psi_{m}(\cdot,\delta)|_{\overline{\Omega_{b}}}$ is continuous in the $C^{1,\gamma}$ norm
for $0<\gamma<1$. This continuity yields the following asymptotic expansions:
 \begin{align}
   (\psi_{M-2}(\cdot,\delta),\phi_{0})_{\Gamma_{h}}=&\psi_{M-2}(\mathbf{o},\delta)\sqrt{h}+\mathcal{O}(h^{3/2}),\label{eq:psiM-2:est}\\
   \|\mathbf{q}\|_{2}=&\mathcal{O}(h^{2}),\label{eq:q:est:2}\\
   \|\nabla\psi_{M-2}(\cdot,\delta)\|_{[L^{2}(R_{h})]}=&\mathcal{O}(h),\label{eq:psiM-2:est:2}\\
   \left(\sum_{m=0}^{M-3}\left\|\frac{1}{n}{\nabla}\psi_{m}\right\|^{2}_{[L^{2}(R_{h})]^{2}}\right)^{1/2}=&\mathcal{O}(h),\label{eq:psiM-2:est:3}   
 \end{align}
 where the constants in the $\mathcal{O}-$terms are independent of $\delta$.
 From these results, the leading behavior of $a_{2,0}$ is:
 \begin{equation}
   \label{eq:a20:asymp}
   a_{2,0}=\psi_{M-2}(\mathbf{o},\delta)\sqrt{h}+\mathcal{O}(h).
 \end{equation}
 An analogous argument gives:
 \begin{equation}
   \label{eq:a21:asymp}
   a_{2,1}=\psi_{M-1}(\mathbf{o},\delta)\sqrt{h}+\mathcal{O}(h).
 \end{equation}
 Condition A.2 implies that
 \begin{equation}
   \label{eq:sec5:ratios:psi}
   \frac{\psi_{M-2}(\mathbf{o},0)}{\psi_{M-1}(\mathbf{o},0)}\ne0\ \text{and}\
   \frac{\psi_{M-1}(\mathbf{o},0)}{\psi_{M-2}(\mathbf{o},0)}\ne0. 
 \end{equation}
 Therefore, we can choose $h_{2},\delta_{3}>0$ such that the ratios
 ${a_{2,0}}/{a_{2,1}}$ and ${a_{2,1}}/{a_{2,0}}$ are uniformly bounded 
 for all $h\in(0,h_{2})$ and $\delta\in[-\delta_{3},\delta_{3}]$.
\end{proof}
\end{theorem}

\subsection{Continuity in $\delta$ and 
Lipschitz continuity in $\mu$ for $f_{0}$ and $f_{1}$}
In the following analysis, we establish the continuity of the
functions $f_{0}(\delta,\mu,h)$ and $f_{1}(\delta,\mu,h)$ in $\delta$, as
well as their Lipschitz continuity in $\mu$ with a Lipschitz constant
less than $1$.
\begin{lemma}
  \label{lem:sec5:Tdeltak}  
  Let $\mathbf{T}(\delta,\mu)$ denote an operator in $\ell^{2}$ that
  \begin{equation}
    \label{eq:sec5:Tdeltak}
    \mathbf{T}(\delta,\mu)=
    \begin{bmatrix}
      \frac{\sqrt{1+\lambda_{M}(\delta)}}{\sqrt{\lambda_{M}(\delta)-\mu}}&&\\
                                                   &\frac{\sqrt{1+\lambda_{M+1}(\delta)}}{\sqrt{\lambda_{M+1}(\delta)-\mu}}&\\
      &&\ddots
    \end{bmatrix}.
  \end{equation}
  Under Condition A.1, $\mathbf{T}(\delta,\mu)$ is continues in $\delta$ and
  $\mu$ for $\delta\in[-\delta_{1},\delta_{1}]$ and $\mu\in[\mu_{l},\mu_{r}]$. Moreover, there is
  a constant $C_{2}$ irrelevant to $\delta$ such that  
  \begin{equation}
    \label{eq:sec5:Tdeltak:est}
    \|\mathbf{T}(\delta,\mu_{a})-\mathbf{T}(\delta,\mu_{b})\|_{\ell^{2}\to\ell^{2}}\le{C_{2}}|\mu_{a}-\mu_{b}|,\
    \text{for}\ \mu_{a},\mu_{b}\in[\mu_{l},\mu_{r}].
  \end{equation}
  \begin{proof}
    Given $m\ge{M}$, each element in
    $\mathbf{T}(\delta_{a},\mu)-\mathbf{T}(\delta_{b},\mu)$ satisfies
    \begin{align}
      &\left|\frac{\sqrt{1+\lambda_{m}(\delta_{a})}}{\sqrt{\lambda_{m}(\delta_{a})-\mu}}-\frac{\sqrt{1+\lambda_{m}(\delta_{b})}}{\sqrt{\lambda_{m}(\delta_{b})-\mu}}\right|\notag\\
      =&\left|\left(\sqrt{1+\frac{\mu+1}{\lambda_{m}(\delta_{a})-\mu}}+\sqrt{1+\frac{\mu+1}{\lambda_{m}(\delta_{b})-\mu}}\right)^{-1}\left(\frac{\mu+1}{\lambda_{m}(\delta_{a})-\mu}-\frac{\mu+1}{\lambda_{m}(\delta_{b})-\mu}\right)\right|\notag\\
      \le&\frac{1}{2}(\lambda_{M}(0)+1)\left(1+\frac{\lambda_{M}(0)+1}{\epsilon}\right)^{2}\left(e^{C_{0}|\delta_{a}-\delta_{b}|}-1\right),\notag        
    \end{align}
    which tends to $0$ uniformly. Then
   the continuity of $\mathbf{T}(\delta,\mu)$ with respect to $\delta$ follows.
   Similarly, each element in
   $\mathbf{T}(\delta,\mu_{a})-\mathbf{T}(\delta,\mu_{b})$ satisfies
    \begin{align}
      &\left|\frac{\sqrt{1+\lambda_{m}(\delta)}}{\sqrt{\lambda_{m}(\delta)-\mu_{a}}}-\frac{\sqrt{1+\lambda_{m}(\delta)}}{\sqrt{\lambda_{m}(\delta)-\mu_{b}}}\right|\notag\\
      =&\sqrt{1+\frac{\mu_{a}+1}{\lambda_{m}(\delta)-\mu_{a}}}\left|\frac{\mu_{a}-\mu_{b}}{\sqrt{\lambda_{m}(\delta)-\mu_{b}}(\sqrt{\lambda_{m}(\delta)-\mu_{a}}+\sqrt{\lambda_{m}(\delta)-\mu_{b}})}\right|\notag\\
      \le&\frac{1}{2\epsilon}\sqrt{1+\frac{\lambda_{M}(0)+1}{\epsilon}}|\mu_{a}-\mu_{b}|,\notag
    \end{align}
   which gives \eqref{eq:sec5:Tdeltak:est}. 
  \end{proof}
\end{lemma}

To study the continuity of the operator
$(\mathbf{I}-\mathbf{B})^{-1}$ in $\ell^{2}$
with respect to $\delta$, we say an operator $\mathbf{G}(\delta)$ in $\ell^{2}$ is
strongly continuous on $[-\delta_{0},\delta_{0}]$ if for every $\mathbf{d}\in\ell^{2}$
\begin{equation}
  \label{eq:sec5:strongconverge}
  \left\|\mathbf{G}(\delta_{a})\mathbf{d}-\mathbf{G}(\delta_{b})\mathbf{d}\right\|_{\ell^{2}}\to0\
  \text{as}\ \delta_{a}\to\delta_{b},\ \text{for}\ \delta_{a},\delta_{b}\in[-\delta_{0},\delta_{0}],
\end{equation}
as defined in \cite[p. 152]{kato95}.

\begin{lemma}
  \label{lem:sec5:Ginverse}  
  Let $\mathbf{G}(\delta)$ denote an invertible, self-adjoint operator in
  $\ell^{2}$ that is strongly continuous with respect to
  $\delta\in[-\delta_{0},\delta_{0}]$. If its inverse $\mathbf{G}^{-1}(\delta)$ is
  uniformly bounded on $[-\delta_{0},\delta_{0}]$, then $\mathbf{G}^{-1}(\delta)$ is
  also strongly continuous.
  \begin{proof}
    Given any uniform bounded vector $\mathbf{v}(\delta)\in\ell^{2}$ and a
    constant vector $\mathbf{q}\in\ell^{2}$, \eqref{eq:sec5:strongconverge} implies 
    \begin{equation}
      \label{eq:sec5:vdelta}
      (\mathbf{v}(\delta_{a}),\mathbf{G}(\delta_{a})\mathbf{q}-\mathbf{G}(\delta_{b})\mathbf{q})_{\ell^{2}}\to0\
      \text{as}\ \delta_{a}\to\delta_{b}.
    \end{equation}
    Define
    $\widetilde{\mathbf{q}}:=\mathbf{G}(\delta_{b})\mathbf{q}$ and
    $\mathbf{v}(\delta_{a}):=\mathbf{G}^{-1}(\delta_{a})(\mathbf{G}^{-1}(\delta_{b})\widetilde{\mathbf{q}}-\mathbf{G}^{-1}(\delta_{a})\widetilde{\mathbf{q}})$.
    Substituting into \eqref{eq:sec5:vdelta} and using the
    self-adjointness of $\mathbf{G}(\delta_{a})$ and uniform boundedness of
    $\mathbf{G}^{-1}(\delta_{a})$, we obtain  
    \begin{equation}
      \label{eq:sec5:Ginverse}
      (\mathbf{G}^{-1}(\delta_{b})\widetilde{\mathbf{q}}-\mathbf{G}^{-1}(\delta_{a})\widetilde{\mathbf{q}},\mathbf{G}^{-1}(\delta_{b})\widetilde{\mathbf{q}}-\mathbf{G}^{-1}(\delta_{a})\widetilde{\mathbf{q}})_{\ell^{2}}\to0\ \text{as}\ \delta_{a}\to\delta_{b},
    \end{equation}
    which implies $\mathbf{G}^{-1}(\delta)$ is strongly continuous.
  \end{proof}
\end{lemma}

\begin{lemma}
  \label{lem:sec5:Hdeltak}  
Let $\mathbf{H}(\delta,\mu)$ be an operator in $\ell^{2}$ defined by
\begin{equation}
\label{eq:sec5:Hdeltak}  
  \mathbf{H}(\delta,\mu):=
  \sum_{j=1}^{\infty}\bi\alpha_{j}(k)\mathbf{w}_{j}(\delta)\mathbf{w}_{j}^{*}(\delta),
\end{equation}
where the vectors $\mathbf{w}_{j}(\delta)$ are given by
\begin{equation}
 \label{eq:sec5:widelta}  
  \mathbf{w}_{j}(\delta):=
  \begin{bmatrix}
    (\phi_{j},\frac{\psi_{M}(\cdot,\delta)}{\sqrt{1+\lambda_{M}(\delta)}})_{\Gamma_{h}}&(\phi_{j},\frac{\psi_{M+1}(\cdot,\delta)}{\sqrt{1+\lambda_{M+1}(\delta)}})_{\Gamma_{h}}&\cdots
  \end{bmatrix}^{T}.   
\end{equation}
Under Condition A.1 and for $h\in(0,h_{0})$,
the operator $\mathbf{H}(\delta,\mu)$ is strongly continuous in $\delta$ for
$\delta\in[-\delta_{2},\delta_{2}]$ and $\mu\in[\mu_{l},\mu_{r}]$.
Moreover, it satisfies the Lipschitz estimate
  \begin{equation}
    \label{eq:sec5:Hdeltak:est}
    \|\mathbf{H}(\delta,\mu_{a})-\mathbf{H}(\delta,\mu_{b})\|_{\ell^{2}\to\ell^{2}}\le{C_{3}}|\mu_{a}-\mu_{b}|\
    \text{for}\ \mu_{a},\mu_{b}\in[\mu_{l},\mu_{r}],
  \end{equation}
  where $C_{3}$ is a constant independent of $\delta$.
  \begin{proof}
    Consider the operator
    \begin{equation}
      \label{eq:sec5:tildeH}      
      \widetilde{\mathbf{H}}(\delta,\mu):=\sum_{j=1}^{\infty}\bi\alpha_{j}(\mu)\widetilde{\mathbf{w}}_{j}(\delta)\widetilde{\mathbf{w}}_{j}^{*}(\delta),
    \end{equation}
    where the vectors $\widetilde{\mathbf{w}}_{j}(\delta)$ are defined by
    \begin{equation}
  \label{eq:sec5:tildewidelta}            
      \widetilde{\mathbf{w}}_{j}(\delta):=  \begin{bmatrix}
    (\phi_{j},\frac{\psi_{0}(\cdot,\delta)}{\sqrt{1+\lambda_{0}(\delta)}})_{\Gamma_{h}}&(\phi_{j},\frac{\psi_{1}(\cdot,\delta)}{\sqrt{1+\lambda_{1}(\delta)}})_{\Gamma_{h}}&\cdots
                                    \end{bmatrix}^{T}.
                                  \end{equation}
The operator $\mathbf{H}(\delta,\mu)$ is then a restriction of
$\widetilde{\mathbf{H}}(\delta,\mu)$. Let $\ell^{2}_{c}$ be a dense subspace of
$\ell^{2}$ consisting of sequences with finite support:
\begin{equation}
  \label{eq:sec5:ellc}
  \ell^{2}_{c}:=\left\{\{\widetilde{d}_{m}\}\in\ell^{2}: \exists{m_{0}\in\mathbb{N}}\ s.t.\ d_{m}=0\
  \text{if}\ m{\ge}m_{0}\right\}.
\end{equation}
Letting $\{\widetilde{d}_{0,m}\}\in\ell^{2}_{c}$, we prove
the continuity of 
$\mathbf{d}(\delta):=\widetilde{\mathbf{H}}(\delta,\mu)\{\widetilde{d}_{0,m}\}$ in
$\ell^{2}$. For any $\{\widetilde{d}_{1,m}\}\in\ell^{2}_{c}$, Lemma
\ref{lem:sec3:eigen:continuity} implies that the inner product 
$(\mathbf{d}(\delta_{a}),\{\widetilde{d}_{1,m}\})_{\ell^{2}}$ converges to
$(\mathbf{d}(\delta_{b}),\{\widetilde{d}_{1,m}\})_{\ell^{2}}$ as $\delta_{a}\to\delta_{b}$
for $\delta_{a},\delta_{b}\in[-\delta_{2},\delta_{2}]$.
By the Banach-Steinhaus Theorem (cf. Theorem 2.7 in \cite{rudin91}),
it follows that $\mathbf{d}(\delta_{a})$ converges weakly to
$\mathbf{d}(\delta_{b})$. Now, let
$\{d_{0,m}\}\in\ell^{2}$ with $\|\{d_{0,m}\}\|_{\ell^{2}}=1$ such that
\begin{align}
  \|\mathbf{d}(\delta_{b})\|_{\ell^{2}}&=(\mathbf{d}(\delta_{b}),\{d_{0,m}\})_{\ell^{2}}\notag\\
  &=\sum_{j=1}^{\infty}\bi\alpha_{j}(\mu)(\phi_{j},\sum_{m=0}^{\infty}\frac{\widetilde{d}_{0,m}\psi_{m}(\cdot,\delta_{b})}{\sqrt{1+\lambda_{m}(\delta_{b})}})_{\Gamma_{h}}(\sum_{m=0}^{\infty}\frac{d_{0,m}\psi_{m}(\cdot,\delta_{b})}{\sqrt{1+\lambda_{m}(\delta_{b})}},\phi_{j})_{\Gamma_{h}}.  \label{eq:sec5:cdelta:norm}
\end{align}
Since $\{{\psi_{m}(\cdot,\delta)}/{\sqrt{1+\lambda_{m}(\delta)}}\}$ forms a complete 
orthonormal basis in $H^{1}(\Omega_{\mathrm{in}})$, there exists a sequence
$\{d_{1,m}\}\in\ell^{2}$ with $\|\{d_{1,m}\}\|_{\ell^{2}}=1$ such that
\begin{equation}
  \label{eq:sec5:cpp}
  \sum_{m=0}^{\infty}\frac{d_{0,m}\psi_{m}(\cdot,\delta_{b})}{\sqrt{1+\lambda_{m}(\delta_{b})}}\Bigg|_{\Gamma_{h}}=\sum_{m=0}^{\infty}\frac{d_{1,m}\psi_{m}(\cdot,\delta_{a})}{\sqrt{1+\lambda_{m}(\delta_{a})}}\Bigg|_{\Gamma_{h}}.
\end{equation}
Consequently, we have the estimate
\begin{align}
  \|\mathbf{d}(\delta_{b})\|_{\ell^{2}}-\|\mathbf{d}(\delta_{a})\|_{\ell^{2}}&\le\|\mathbf{d}(\delta_{b})\|_{\ell^{2}}-(\mathbf{d}({\delta_{a}}),\{d_{1,m}\})_{\ell^{2}}\notag\\
  {\le}n_{1}^{2}&\left\|\frac{1}{n(\cdot,\delta_{a})}\nabla\left(\sum_{m=0}^{\infty}\frac{\widetilde{d}_{0,m}\psi_{m}(\cdot,\delta_{b})}{\sqrt{1+\lambda_{m}(\delta_{b})}}-\sum_{m=0}^{\infty}\frac{\widetilde{d}_{0,m}\psi_{m}(\cdot,\delta_{a})}{\sqrt{1+\lambda_{m}(\delta_{a})}}\right)\right\|_{[L^{2}(R_{h})]^{2}}\label{eq:sec5:strongconvergence}
\end{align}
where the second inequality follows from Lemma
\ref{lem:sec4:estimate}.
Since $\{\widetilde{d}_{0,m}\}\in\ell^{2}_{c}$, Lemma
\ref{lem:sec3:eigen:continuity} implies that the right-hand side of
\eqref{eq:sec5:strongconvergence} vanishes as
$\delta_{a}\to\delta_{b}$. Exchanging $\delta_{a}$ and $\delta_{b}$ shows that
\begin{equation}
  \label{eq:sec5:converge:norm}
  \|\mathbf{d}(\delta_{a})\|_{\ell^{2}}\to\|\mathbf{d}(\delta_{b})\|_{\ell^{2}},\ \text{as}\ \delta_{a}\to\delta_{b}.
\end{equation}
This convergence of norms, together with the established weak convergence,
implies the $\ell^{2}$ convergence of
$\mathbf{d}(\delta_{a})$ to $\mathbf{d}(\delta_{b})$ (cf. Exercise 2.11 in
\cite{mcl00}). Therefore $\widetilde{\mathbf{H}}(\delta,\mu)$ is strongly
continuous by the Banach-Steinhaus Theorem.

  We now prove the Lipschitz continuity in $\mu$, i.e.,
  \eqref{eq:sec5:Hdeltak:est}. For any
  $\{d_{0,m}\},\{d_{1,m}\}\in\ell^{2}$, we have
  \begin{align}
    &\left|((\widetilde{\mathbf{H}}(\delta,\mu_{a})-\widetilde{\mathbf{H}}(\delta,\mu_{b}))\{d_{0,m}\},\{d_{1,m}\})_{\ell^{2}}\right|\notag\\
    =&\sum_{j=1}^{\infty}\left|\bi(\alpha_{j}(\mu_{a})-\alpha_{j}(\mu_{b}))\right|(\phi_{j},\sum_{m=0}^{\infty}\frac{d_{0,m}\psi_{m}(\cdot,\delta)}{\sqrt{1+\lambda_{m}(\delta)}})_{\Gamma_{h}}(\sum_{m=0}^{\infty}\frac{d_{1,m}\psi_{m}(\cdot,\delta)}{\sqrt{1+\lambda_{m}(\delta)}},\phi_{j})_{\Gamma_{h}}\notag\\
    \le&\frac{1}{2\epsilon^{2}}|\mu_{a}-\mu_{b}|\|\{d_{0,m}\}\|_{\ell^{2}}\|\{d_{1,m}\}\|_{\ell^{2}},\
       \text{for}\ \mu_{a},\mu_{b}\in[\mu_{l},\mu_{r}].
  \end{align}
  The inequality relies on Lemma
  \ref{lem:sec4:estimate} and the bound
  \begin{equation}
    \label{eq:sec5:alpha:est}
    |\bi(\alpha_{j}(\mu_{a})-\alpha_{j}(\mu_{b}))|\le\frac{\bi\alpha_{j}(\mu_{b})|\mu_{b}-\mu_{a}|}{(\bi\alpha_{j}(\mu_{a})+\bi\alpha_{j}(\mu_{b}))\bi\alpha_{j}(\mu_{b})}\le\frac{\bi\alpha_{j}(\mu_{b})|\mu_{b}-\mu_{a}|}{2\epsilon^{2}}.
  \end{equation}
  \end{proof}
\end{lemma}

\begin{theorem}
  \label{thm:sec5:continuity:Lipcontinuity}
  Under Conditions A.1, A.2, there exists $h_{3}>0$ such that for all
  $h\in(0,h_{3})$, $\delta\in[-\delta_{3},\delta_{3}]$ and $\mu\in[\mu_{l},\mu_{r}]$,
  the functions $f_{0}(\delta,\mu,h)$ and $f_{1}(\delta,\mu,h)$ are continuous in
  $\delta$ and Lipschitz continuous in $\mu$. Moreover, the following
  Lipschitz estimates hold for all $\mu_{a},\mu_{b}\in[\mu_{l},\mu_{r}]$:  
\begin{align}
  |f_{0}(\delta,\mu_{a},h)-f_{0}(\delta,\mu_{b},h)|\le&{C_{f_{0}}}|\mu_{a}-\mu_{b}|,\label{eq:sec5:f0:lipcontinuity}\\
  |f_{1}(\delta,\mu_{a},h)-f_{1}(\delta,\mu_{b},h)|\le&{C_{f_{1}}}|\mu_{a}-\mu_{b}|,\label{eq:sec5:f1:lipcontinuity}
\end{align}
where the constants $C_{f_{0}}<1$ and $C_{f_{1}}<1$ are independent of
$\delta$ and $h$.
\begin{proof}
  By neglecting the diagonal matrix
  $\mathrm{Diag}(\{\lambda_{m}-\mu\}_{m=0}^{M-1})$, we will show that
  each entry of the finite-dimensional system  
  \eqref{eq:sec4:FLS:real}--\eqref{eq:sec4:FLS:imag} is continuous in
  $\delta$ and Lipschitz continuous in $\mu$, with a Lipschitz constant of
  order $\mathcal{O}(h)$. The desired result follows from this.
  Consider the $(m_{1},m_{2})-$entry of $\mathbf{A}$ for
  $0{\le}m_{1},m_{2}\le{M-1}$. Its difference for two values of $\delta$ is
  bounded by
  \begin{align}
    &\Bigg|\sum_{j=1}^{\infty}\bi\alpha_{j}(\mu)(\phi_{j},\psi_{m_{1}}(\cdot,\delta_{a}))_{\Gamma_{h}}(\psi_{m_{2}}(\cdot,\delta_{a}),\phi_{j})_{\Gamma_{h}}\notag\\
    &\qquad\qquad-\sum_{j=1}^{\infty}\bi\alpha_{j}(\mu)(\phi_{j},\psi_{m_{1}}(\cdot,\delta_{b}))_{\Gamma_{h}}(\psi_{m_{2}}(\cdot,\delta_{b}),\phi_{j})_{\Gamma_{h}}\Bigg|\notag\\
    \le&n_{1}^{2}\|\frac{1}{n(\cdot,\delta_{a})}\nabla(\psi_{m_{1}}(\cdot,\delta_{a})-\psi_{m_{1}}(\cdot,\delta_{b}))\|_{[L^{2}(R_{h})]^{2}}\|\frac{1}{n(\cdot,\delta_{a})}\nabla(\psi_{m_{2}}(\cdot,\delta_{a}))\|_{[L^{2}(R_{h})]^{2}}\notag\\
    &+n_{1}^{2}\|\frac{1}{n(\cdot,\delta_{a})}\nabla(\psi_{m_{2}}(\cdot,\delta_{a})-\psi_{m_{2}}(\cdot,\delta_{b}))\|_{[L^{2}(R_{h})]^{2}}\|\frac{1}{n(\cdot,\delta_{b})}\nabla(\psi_{m_{1}}(\cdot,\delta_{b}))\|_{[L^{2}(R_{h})]^{2}}.\notag
  \end{align}
  By Lemma \ref{lem:sec3:eigen:continuity}, this entry is continuity
  in $\delta$. And its Lipschitz continuity in $\mu$ follows from Lemmas
  \ref{lem:sec4:estimate}, 
  \ref{lem:sec5:H3converge} and the general Sobolev inequality,
  which give
  \begin{align}
    \left|\sum_{j=1}^{\infty}\bi(\alpha_{j}(\mu_{a})-\alpha_{j}(\mu_{b}))(\phi_{j},\psi_{m}(\cdot,\delta))_{\Gamma_{h}}(\psi_{m}(\cdot,\delta),\phi_{j})_{\Gamma_{h}}\right|\le\mathcal{O}(h^{2})|\mu_{a}-\mu_{b}|,
  \end{align}
  where the constant hided in $\mathcal{O}$ is independent of $\delta$. Let the vector
  $\mathbf{z}_{m_{1}}(\delta,\mu)$ be defined by  
  \begin{equation}
    \label{eq:sec5:zmdeltak}
    \mathbf{z}_{m_{1}}(\delta,\mu):=\sum_{j=1}^{\infty}\bi\alpha_{j}(\mu)(\phi_{j},\psi_{m_{1}}(\cdot,\delta))_{\Gamma_{h}}\mathbf{w}_{j}(\delta).
  \end{equation}
  The scaled vector $\mathbf{z}_{m_{1}}(\delta,\mu)/\sqrt{1+\lambda_{m_{1}}(\delta)}$ is a
  restriction of the $m_{1}$-th column of the operator
  $\widetilde{\mathbf{H}}(\delta,\mu)$ from \eqref{eq:sec5:tildeH}. Applying
  Lemma \ref{lem:sec5:H3converge} and the general Sobolev inequality
  yields the estimates:
  \begin{align}
    \|\mathbf{z}_{m_{1}}(\delta,\mu)\|_{\ell^{2}}&=\mathcal{O}(h),    \label{eq:sec5:zmdeltak:h}\\
    \|\mathbf{z}_{m_1}(\delta, \mu_1) - \mathbf{z}_{m_1}(\delta, \mu_2)\|_{\ell^{2}}&=\mathcal{O}(h)|\mu_1 - \mu_2|.    \label{eq:sec5:zmdeltak:liph}
  \end{align}
  Using techniques analogous to those in Lemma \ref{lem:sec5:Hdeltak},
  we also find that the vector $\mathbf{w}_{0}(\delta)$ is continuous
  in $\delta$. The proof is completed by observing that
  the $(m_{1},m_{2})$-entry of
  $\mathbf{V}^{*}(\mathbf{I}-\mathbf{B})^{-1}\mathbf{V}$ admits  
  the representation
  $\mathbf{z}_{m_{1}}^{*}\mathbf{T}(\mathbf{I}-\mathbf{T}\mathbf{H}\mathbf{T})^{-1}\mathbf{T}\mathbf{z}_{m_{2}}$,
  and the $m_{2}$-th column entry of
  $\mathbf{v}^{*}_{0,M-1,0}\mathbf{D}(\mathbf{I}-\mathbf{B})^{-1}\mathbf{V}$
  can be written as
  $\mathbf{w}_{0}^{*}\mathbf{T}(\mathbf{I}-\mathbf{T}\mathbf{H}\mathbf{T})^{-1}\mathbf{T}\mathbf{z}_{m_{2}}$. Their
  continuity in $\delta$ and Lipschitz continuity in $\mu$ follow from Lemmas
  \ref{lem:sec5:Tdeltak}, \ref{lem:sec5:Ginverse}, and
  \ref{lem:sec5:Hdeltak}. And the value $h_{3}$ is chosen such that 
  \eqref{eq:sec5:f0:lipcontinuity}--\eqref{eq:sec5:f1:lipcontinuity}
  hold for all $h\in(0,h_{3})$.
\end{proof}
\end{theorem}

\subsection{Existence of Friedrich-Wintgen BICs}
In the following theorem, we prove that each equation
\eqref{eq:sec4:exBIC:cond1}--\eqref{eq:sec4:exBIC:cond2} defines an implicit function of $\mu(\delta)$.
Consequently, BICs correspond to the intersections of the graphs of these
functions.
\begin{theorem}
  \label{thm:sec5:BIC:existence}
  {Under Conditions A.1--A.3, there exists $h_{4}>0$ such that for every
  $h\in(0,h_{4})$, the system \eqref{eq:sec2:goveq}--\eqref{eq:sec2:bc} admits
  a BIC for some $\delta\in[-\delta_{3},\delta_{3}]$ and
  $\mu\in[\mu_{l},\mu_{r}]$, where $\delta$ and $\mu$ depend on $h$.}
  \begin{proof}
    Let $\delta_{l}:=-\delta_{3}$ and $\delta_{r}:=\delta_{3}$. By Theorem
    \ref{thm:sec5:f0f1}, for sufficiently small $h$, we have
    \begin{equation}
      \label{eq:sec5:f0f1:epsilon}
    |f_{0}(\delta,\mu,h)|<\epsilon\ \text{and}\ |f_{1}(\delta,\mu,h)|<\epsilon\ \text{for}\
    \delta\in[\delta_{l},\delta_{r}],\ \mu\in[\mu_{l},\mu_{r}].
    \end{equation}
    For a fixed $\delta$, this implies the inequalities:
    \begin{align}
      \lambda_{M-2}(\delta)-\mu_{l}+f_{0}(\delta,\mu_{l},h)>0,\label{eq:sec5:kl>0}\\
      \lambda_{M-2}(\delta)-\mu_{r}+f_{0}(\delta,\mu_{r},h)<0.\label{eq:sec5:kr>0}
    \end{align}
    Since $f_{0}$ is continuous in $\mu$ by Theorem
    \ref{thm:sec5:continuity:Lipcontinuity}, 
    the Intermediate Value Theorem guarantees the existence of
    $\mu_{0}(\delta)\in(\mu_{l},\mu_{r})$ satisfying    
    \begin{equation}
      \label{eq:sec5:kdelta}
      \lambda_{M-2}(\delta)-\mu_{0}(\delta)+f_{0}(\delta,\mu_{0}(\delta),h)=0.
    \end{equation}
    And the Lipschitz condition \eqref{eq:sec5:f0:lipcontinuity}
    ensures $\mu_{0}$ is uniquely defined. The continuity of
    $\mu_{0}(\delta)$ follows from estimating the difference for
    $\delta_{a},\delta_{b}\in[\delta_{l},\delta_{r}]$:
    \begin{align}
      |\mu_{0}(\delta_{a})-\mu_{0}(\delta_{b})|=&|\lambda_{M-2}(\delta_{a})-\lambda_{M-2}(\delta_{b})+f_{0}(\delta_{a},\mu_{0}(\delta_{a}),h)-f_{0}(\delta_{b},\mu_{0}(\delta_{b}),h)|\notag\\
      \le&|\lambda_{M-2}(\delta_{a})-\lambda_{M-2}(\delta_{b})|+|f_{0}(\delta_{a},\mu_{0}(\delta_{b}),h)-f_{0}(\delta_{b},\mu_{0}(\delta_{b}),h)|\notag\\
                                  &+|f_{0}(\delta_{a},\mu_{0}(\delta_{a}),h)-f_{0}(\delta_{a},\mu_{0}(\delta_{b}),h)|\notag\\
      \le&|\lambda_{M-2}(\delta_{a})-\lambda_{M-2}(\delta_{b})|+|f_{0}(\delta_{a},\mu_{0}(\delta_{b}),h)-f_{0}(\delta_{b},\mu_{0}(\delta_{b}),h)|\notag\\
         &+C_{f_{0}}|\mu_{0}(\delta_{a})-\mu_{0}(\delta_{b})|.\label{eq:sec5:kdelta:continuity}
    \end{align}
    Since $C_{f_{0}}<1$, this inequality implies the continuity of
    $\mu_{0}(\delta)$. A similar argument shows the existence of a unique,
    continuous function $\mu_{a}(\delta)$ satisfying
  \begin{equation}
      \label{eq:sec5:kdelta:2}
      \lambda_{M-1}(\delta)-\mu_{a}(\delta)+f_{1}(\delta,\mu_{a}(\delta),h)=0.
    \end{equation}
    Invoking Theorem \ref{thm:sec5:f0f1}, we further take $h$ sufficiently
    small to satisfy
  \begin{align}
    &|f_{0}(\cdot,\cdot,h)|+|f_{1}(\cdot,\cdot,h)|\notag\\
    \le&\min\Big\{|\max_{\delta\in[\delta_{l},\delta_{r}]}\left\{\lambda_{M-2}(\delta)-\lambda_{M-1}(\delta)\right\}|,\
       |\min_{\delta\in[\delta_{l},\delta_{r}]}\left\{\lambda_{M-2}(\delta)-\lambda_{M-1}(\delta)\right\}|\Big\}.    \label{eq:sec5:f0f1h:est}
  \end{align}
  From this and Condition A.3, it follows that
  \begin{equation}
    \label{eq:sec5:changesignmu}
    \max_{\delta\in[\delta_{l},\delta_{r}]}\{\mu_{0}(\delta)-\mu_{1}(\delta)\}\ge0\ \text{and}\  
    \min_{\delta\in[\delta_{l},\delta_{r}]}\{\mu_{0}(\delta)-\mu_{1}(\delta)\}\le0.
  \end{equation}
  Therefore, the continuous function $\mu_{0}(\delta)-\mu_{1}(\delta)$ must have a
  zero in $[\delta_{l},\delta_{r}]$  
  by the Intermediate Value Theorem. We may thus choose $h_{4}>0$
  such that \eqref{eq:sec5:f0f1:epsilon} and
  \eqref{eq:sec5:f0f1h:est} are satisfied for all $h\in(0,h_{4})$.
  \end{proof}
\end{theorem}

\begin{remark}
  \label{rmk:sec5:necessity}
In our analysis, Condition A.1 ensures the continuous dependence
of eigenvalues and eigenfunctions on the parameter $\delta$ for the
boundary value problem
\eqref{eq:sec2:eigen:goveq}--\eqref{eq:sec2:eigen:bc}. Specifically,
the analytic dependence of the refractive index $n(\cdot,\delta)$ on $\delta$
implies:
\begin{itemize}
  \item The continuity of eigenvalues
    \eqref{eq:sec3:continuity:eigenvalue},
  \item The $H^{1}(\Omega_{\mathrm{in}})$-continuity of eigenfunctions \eqref{eq:sec3:H1converge}.
\end{itemize}
These results enable the reduction of the infinite-dimensional
system \eqref{eq:sec3:ILS} to a finite-dimensional system 
\eqref{eq:sec4:simLS} for $\delta$ and $\mu$ over small intervals and $h\ll1$.
By further requiring 
$n(\mathbf{x},\delta)=1$ for $\mathbf{x}\in\Omega_{B}$, we can derive
the uniform boundedness of
the eigenfunctions and their derivatives on $\Gamma_{h}$ and
its neighboring regions as $\delta$ varies.
Combining Condition A.1 with
Condition A.2, the two governing equations
\eqref{eq:sec4:exBIC:cond1}--\eqref{eq:sec4:exBIC:cond2} are
well-defined. Furthermore, these conditions guarantee the
following critical properties for the functions $f_{0}(\delta,\mu,h)$ and
$f_{1}(\delta,\mu,h)$ defined in \eqref{eq:sec5:f0}--\eqref{eq:sec5:f1} for $h\ll1$:
\begin{itemize}
    \item uniform convergence: $f_{0},f_{1}\to0$ as $h\to0$ uniformly;
    \item continuity in $\delta$: $f_{0}$ and $f_{1}$ are continuous with
      respect to $\delta$;
    \item Lipschitz continuity in $\mu$: both $f_{0}$ and $f_{1}$ are
      Lipschitz continuous in $\mu$ with a Lipschitz constant less than 1.
\end{itemize}
Consequently, each governing equations
\eqref{eq:sec4:exBIC:cond1}--\eqref{eq:sec4:exBIC:cond2} defines an
implicit function $\mu(\delta)$. BICs emerge as intersections of these
implicit functions. Under Condition A.3, the existence of such an
intersection is rigorously guaranteed.
\end{remark}

\section{Numerical experiments}
  In \cite{lyapina15}, the emergence of
  FW-BICs in an acoustic system was numerically demonstrated for
  homogeneous rectangular  
  cavities with waveguide openings. If the
  cavity supports a multiple  
  eigenvalue, a FW-BIC can be achieved by perturbing the cavity length
  slightly. 
  Here, we consider an inhomogeneous, geometrically asymmetric region,
  governed by the PDE system \eqref{eq:sec2:goveq}--\eqref{eq:sec2:bc}.
  Through numerical experiments, we show that a FW-BIC can be
  achieved  
  by introducing a slight perturbation
  to the refractive index $n(x)$, assuming the closed cavity has a
  multiple eigenvalue when there is no perturbation.

  \subsection{Example 1}
\begin{figure}[htbp]
    \centering
  \subfigure[]{
    \centering
        \includegraphics[width=0.2\textwidth]{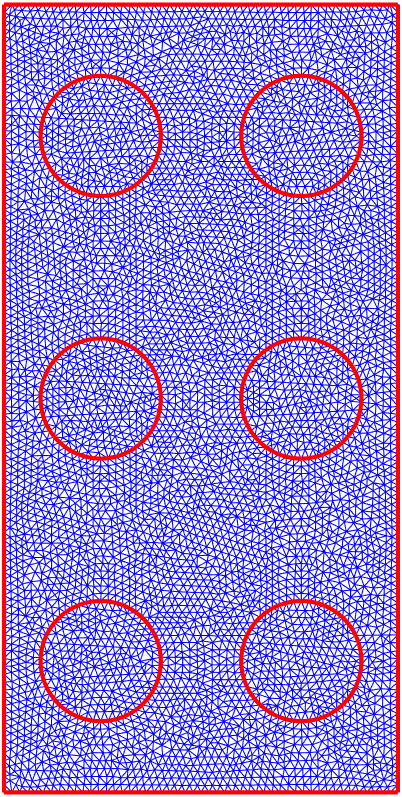}
    \label{fig:sec6:mesh0}}\qquad\qquad\qquad
  \subfigure[]{    
    \centering
        \includegraphics[width=0.595\textwidth]{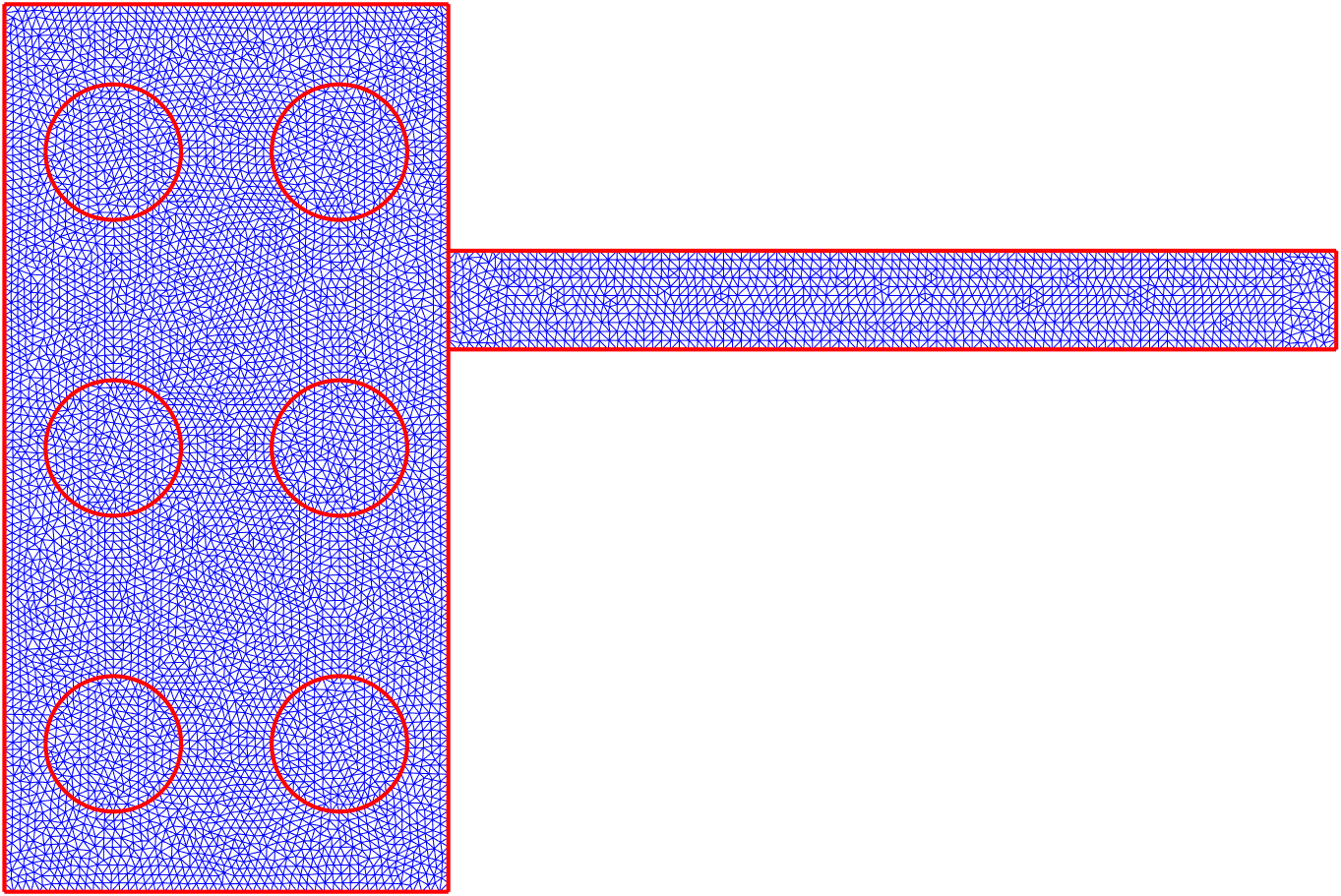}   
    \label{fig:sec6:mesh1}}  
  \caption{(a). Computational mesh used for the eigenvalue problem
    \eqref{eq:sec2:eigen:goveq}--\eqref{eq:sec2:eigen:bc} in    
    $\Omega_{\mathrm{in}}$. (b) Computational mesh used for computing the
    resonances of the system governed by
    \eqref{eq:sec2:goveq}--\eqref{eq:sec2:bc} in $\Omega$ with $h=2\pi/9$.}
\end{figure}

As illustrated in Fig. \ref{fig:sec6:numexp},
the full computational region $\Omega$ is partitioned into three
subregions:
\begin{itemize}
\item $\Omega_{\mathrm{in}}$: the rectangular cavity,
\item $\Omega_{\mathrm{out}}$: a semi-infinite waveguide of width $h=2\pi/9$,
\item $\Gamma_{h}$: the shared boundary between $\Omega_{\mathrm{in}}$ and $\Omega_{\mathrm{out}}$.
\end{itemize}
A rectangular coordinate system is centered at the midpoint $\mathbf{o}$ of $\Gamma_{h}$, with the cavity extending between the coordinates $\mathbf{o}_{1}=(0,-4\pi/3)$ and $\mathbf{o}_{2}=(-\pi,2\pi/3)$.
Six identical circular elements (radius 0.48) are symmetrically embedded within the cavity. Their centers are located at:
\begin{equation}
  \label{eq:sec6:centers}
  (-\pi/2\pm0.8,\pi/3),\ (-\pi/2\pm0.8,-\pi/3)\ \text{and}\ (-\pi/2\pm0.8,-\pi).
\end{equation}
The refractive index $n(\mathbf{x}) $ is modeled as a piecewise
constant function: it assumes the same constant value within each
element and is set to $1$ in the surrounding ambient medium.
We use a Perfectly Matched Layer (PML) to absorb outgoing fields and
employ the piecewise linear continuous finite-element method to
formulate the eigenvalue problems. Fig. \ref{fig:sec6:mesh0} and \ref{fig:sec6:mesh1} show examples of our
computational meshes, all of which are constructed uniformly.
\begin{figure}[htbp]
  \centering
    \begin{tikzpicture}
      \draw [ultra thick](0,pi/9) -- (1.5*pi,pi/9);
      \draw [ultra thick](0,-pi/9) -- (1.5*pi,-pi/9);      
      \draw [ultra thick](0,-pi/9) -- (0,-4*pi/3);
      \draw [ultra thick](0,-4*pi/3) -- (-pi,-4*pi/3);
      \draw [ultra thick](-pi,-4*pi/3) -- (-pi,2*pi/3);
      \draw [ultra thick](-pi,2*pi/3) -- (0,2*pi/3);
      \draw [ultra thick](0,2*pi/3) -- (0,pi/9);
      \draw (0.1,-0.2) node[red]{$\mathbf{o}$};
      \draw [->,thick,red,dashed](0,0) -- (1.5,0) node [right,pos=1] {$x_{1}$};
      \draw [->,thick,red,dashed](0,0) -- (0,1.5) node [right,pos=1]
      {$x_{2}$};
      \fill [lightgray] (-pi/2+0.8,pi/3) circle (0.48);
      \fill [lightgray] (-pi/2-0.8,pi/3) circle (0.48);
      \fill [lightgray] (-pi/2+0.8,-pi/3) circle (0.48);
      \fill [lightgray] (-pi/2-0.8,-pi/3) circle (0.48);
      \fill [lightgray] (-pi/2+0.8,-pi) circle (0.48);
      \fill [lightgray] (-pi/2-0.8,-pi) circle (0.48);      
      
      \draw (5*pi/6,0) node {$\Omega_{\mathrm{out}}$};
      \draw (-pi/2,-4*pi/6) node {$\Omega_{\mathrm{in}}$};
      \draw (-pi/2+0.8,-pi/3) node {\small$n(\mathbf{x})$};
      \draw (0,-4*pi/3) node[right] {$\mathbf{o}_{1}$};
      \draw (-pi,2*pi/3) node[left] {$\mathbf{o}_{2}$};      
      \draw (5*pi/6,0) node {$\Omega_{\mathrm{out}}$};
      \draw (-0.3,0) node {$\Gamma_{h}$};
      \draw [<->,thick](4,-pi/9) -- (4,pi/9) node [right,pos=0.5]
      {$h$};      
\end{tikzpicture}        
    \label{fig:sec6:numexp}
    \caption{Schematic of a rectangular cavity with a waveguide
      opening. The computational region is defined as
      $\Omega:=\Omega_{\mathrm{in}}\cup\Gamma_{h}\cup\Omega_{\mathrm{out}}$ comprises a
      rectangular cavity $\Omega_{\mathrm{in}}$ spanning between
      coordinates $\mathbf{o}_{1}=(0,-4\pi/3)$ and
      $\mathbf{o}_{2}=(-\pi,2\pi/3)$, a semi-infinite waveguide
      $\Omega_{\mathrm{out}}$ of width $h=2\pi/9$, and their shared interface
      $\Gamma_{h}$. A rectangular coordinate system is centered at
      $\mathbf{o}$, the midpoint of $\Gamma_{h}$. Six identical circular
      elements (radius 0.48) with refractive index
      $n(\cdot)$ are symmetrically embedded in $\Omega_{\mathrm{in}}$,
      centered at positions 
      $(-\pi/2\pm0.8,\pi/3)$, $(-\pi/2\pm0.8,-\pi/3)$ and
      $(-\pi/2\pm0.8,-\pi)$.}
  \end{figure}
  
\begin{figure}[htbp]
    \centering
    \includegraphics[width=0.9\textwidth]{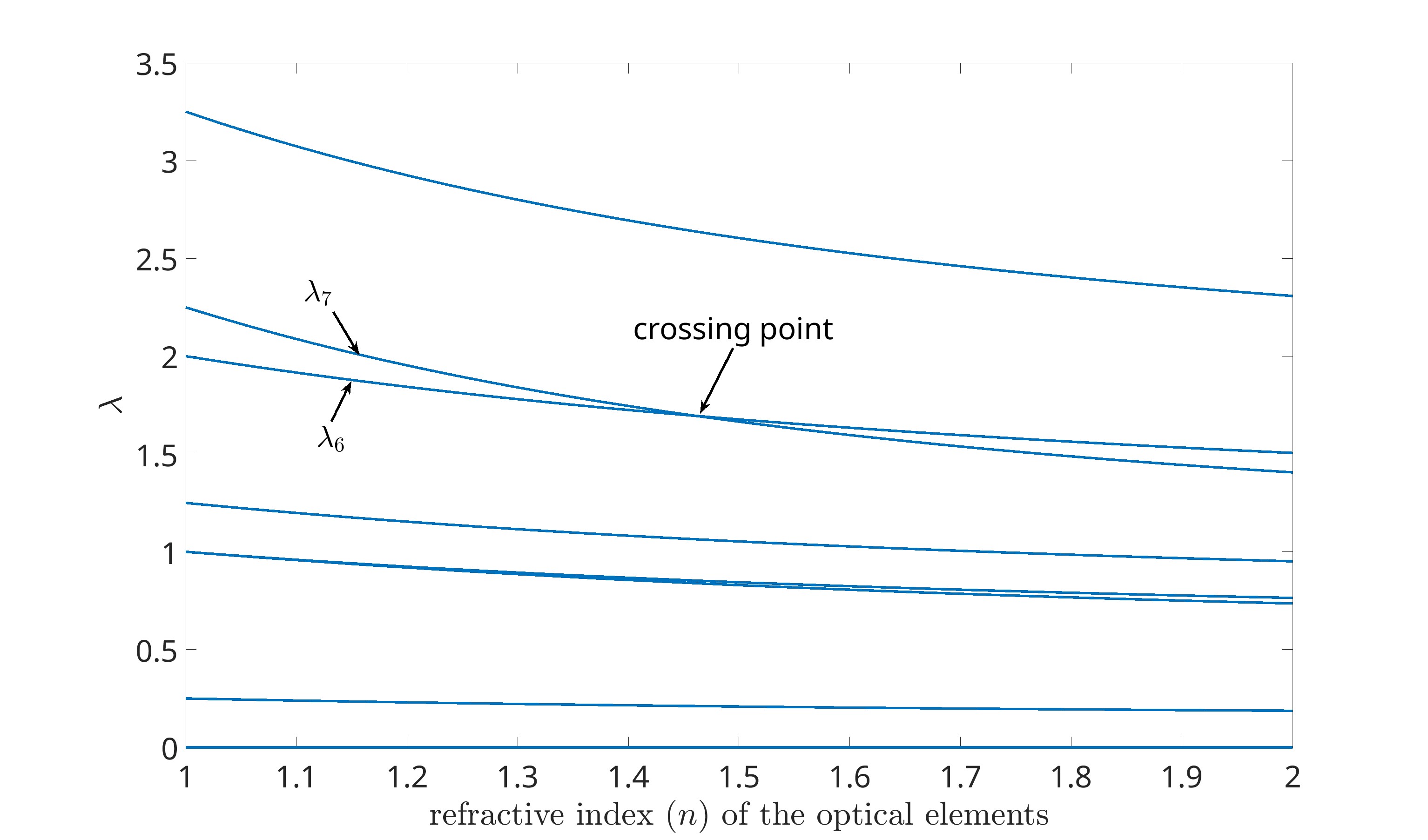}
    \label{fig:sec6:eigenstructure}
    \caption{First eight eigenvalues of the problem
      \eqref{eq:sec2:eigen:goveq}--\eqref{eq:sec2:eigen:bc} in
      $\Omega_{\mathrm{in}}$ as the refractive
      index of the elements varies from $1$ to $2$.}
\end{figure}

For the eigenvalue problem
\eqref{eq:sec2:eigen:goveq}--\eqref{eq:sec2:eigen:bc} in the region
$\Omega_{\mathrm{in}}$, Figure \ref{fig:sec6:eigenstructure} illustrates
the evolution of the first eight eigenvalues as the refractive index
$n$ of the elements increases from $1$ to $2$. Notably, the
sixth and seventh eigenvalues, labeled $\lambda_{6}$ and
  $\lambda_{7}$, {intersect transversally} at
  $n\approx1.461$, resulting in a multiple eigenvalue $\lambda_{6}=\lambda_{7}\approx1.695$.

\begin{figure}[htbp]
  \centering
  \subfigure[Real parts of $\widetilde{\lambda}_{6}$ and $\widetilde{\lambda}_{7}$.]{
    \centering
        \includegraphics[width=0.9\textwidth]{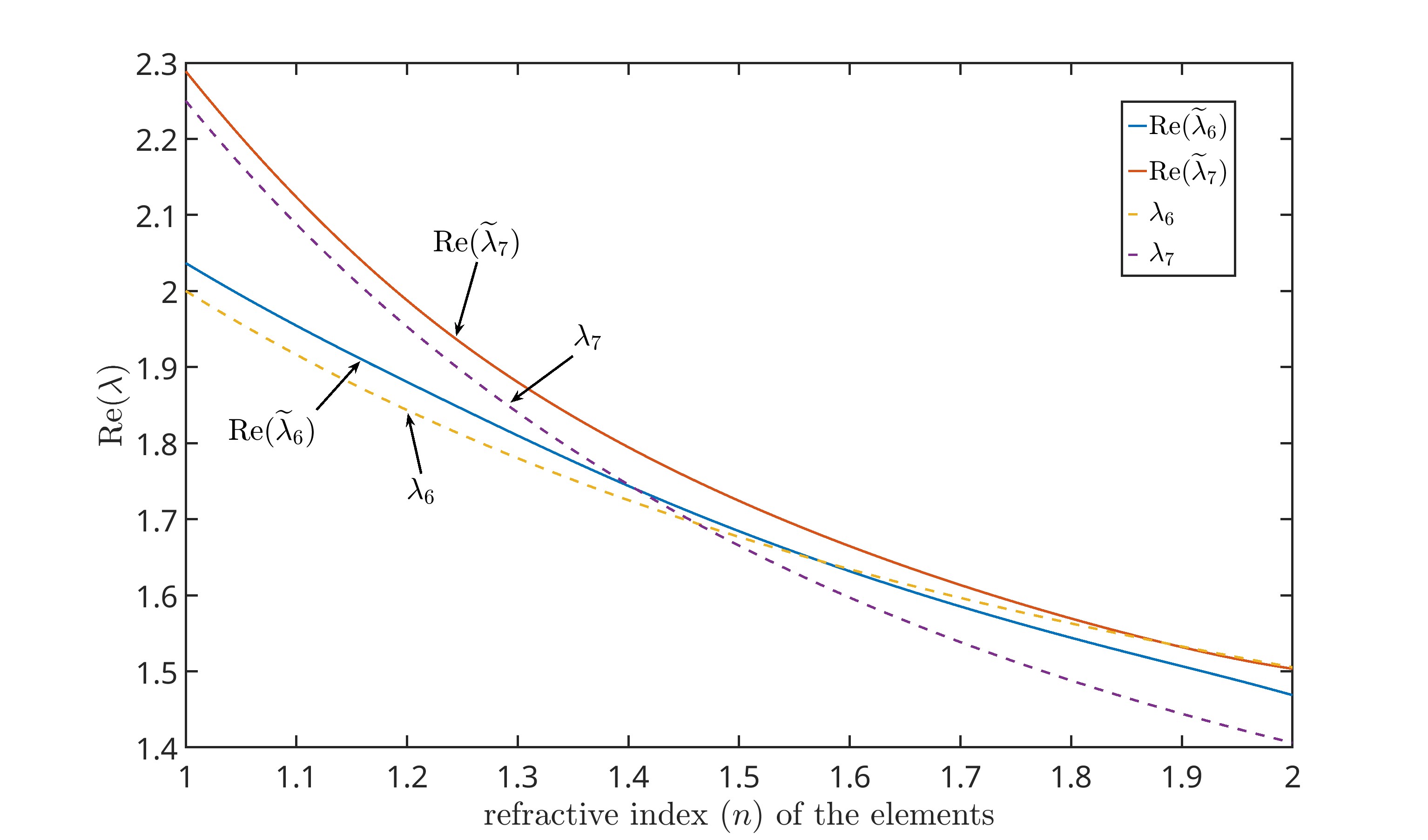}
    \label{fig:sec6:realresonance}}\\
  \subfigure[Imaginary parts of $\widetilde{\lambda}_{6}$ and
  $\widetilde{\lambda}_{7}$.]{    
    \centering
        \includegraphics[width=0.9\textwidth]{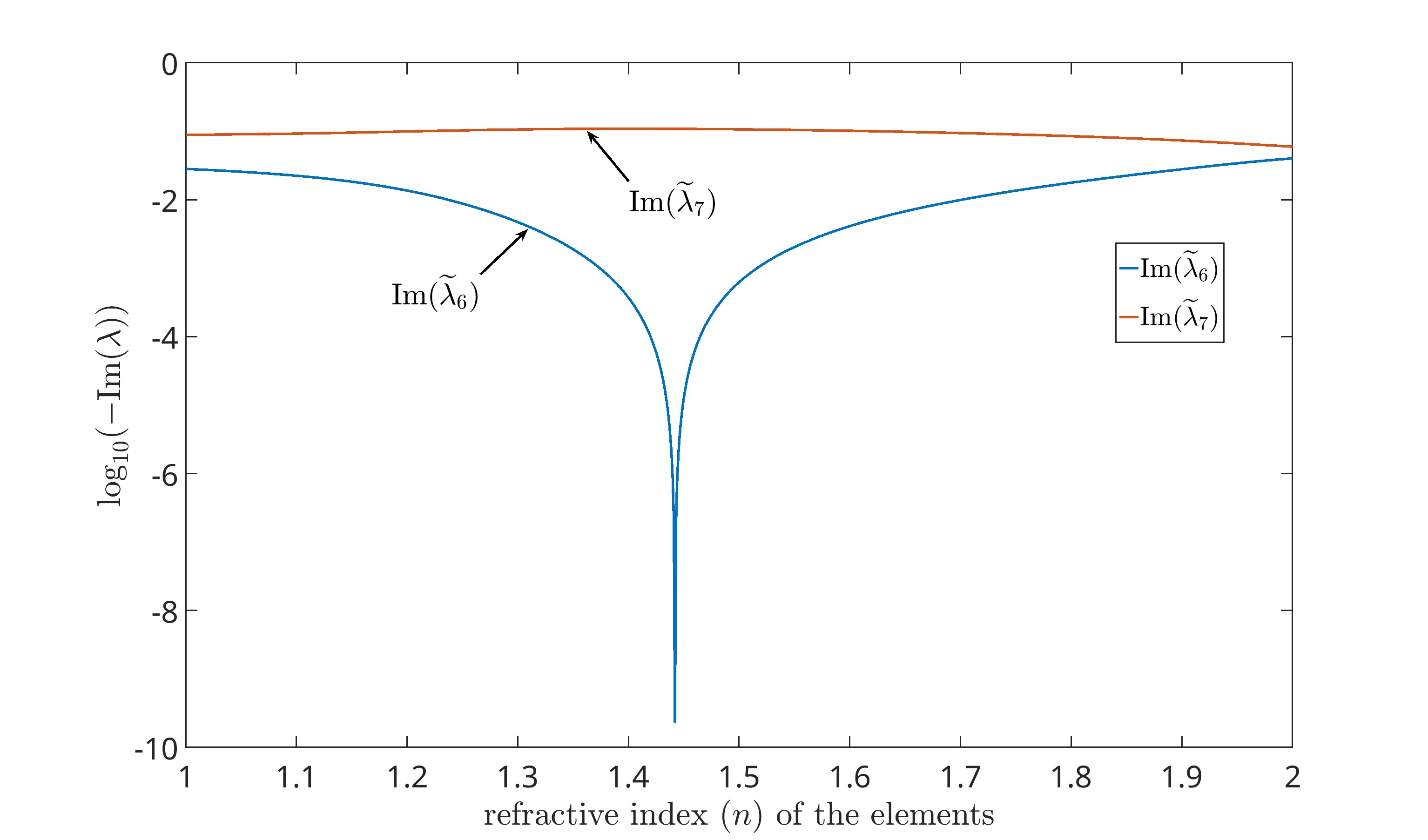}   
        \label{fig:sec6:imagresonance}}\\
   \subfigure[The BIC mode at $n\approx1.442$ and
      $\mu=\widetilde{\lambda}_{6}\approx1.718$.]{        
    \centering
        \includegraphics[width=0.63\textwidth]{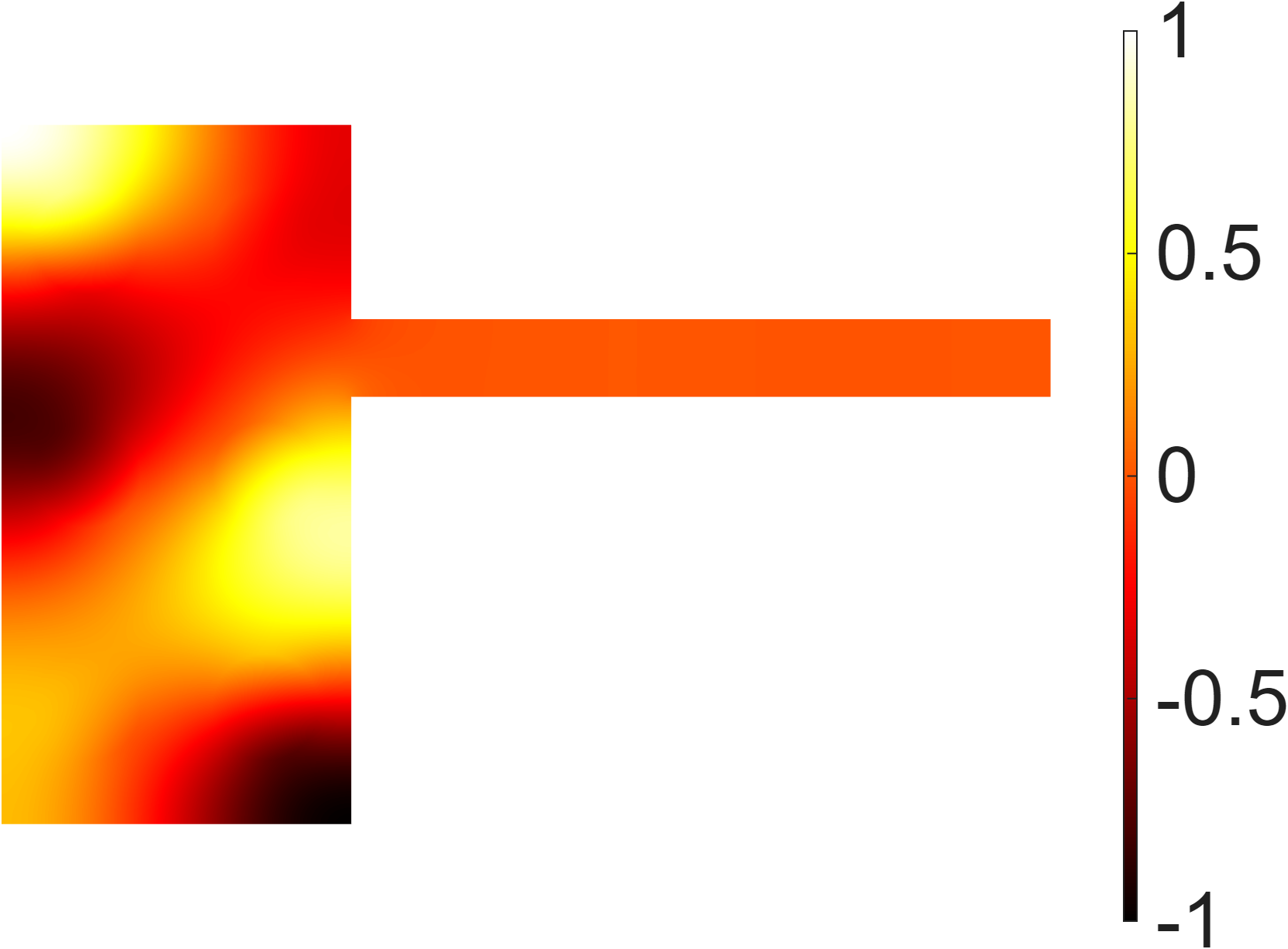}   
    \label{fig:sec6:BIC1}}        
  \caption{Resonances of the problem
    \eqref{eq:sec2:goveq}--\eqref{eq:sec2:bc} in $\Omega$ approximating the
    eigenvalues $\lambda_{6}$ and $\lambda_{7}$ as the refractive index of the
    elements varies. A BIC emerges at $n\approx1.442$ and
      $\mu=\widetilde{\lambda}_{6}\approx1.718$.}  
\end{figure}

When a thin waveguide is introduced on the boundary of
$\Omega_{\mathrm{in}}$, the resulting resonances approximating $\lambda_{6}$ and
$\lambda_{7}$ are denoted by
$\widetilde{\lambda}_{6}$ and $\widetilde{\lambda}_{7}$. Their real
and imaginary
parts are plotted
in Fig. \ref{fig:sec6:realresonance} and
\ref{fig:sec6:imagresonance}, respectively. A
BIC emerges at $n\approx1.442$ and
$\widetilde{\lambda}_{6}\approx1.718$, confirming our theoretical
predictions. The corresponding BIC mode is shown in
Fig. \ref{fig:sec6:BIC1}.
    
\subsection{Example 2} 
\begin{figure}[htbp]
    \centering
    \centering
        \includegraphics[width=0.595\textwidth]{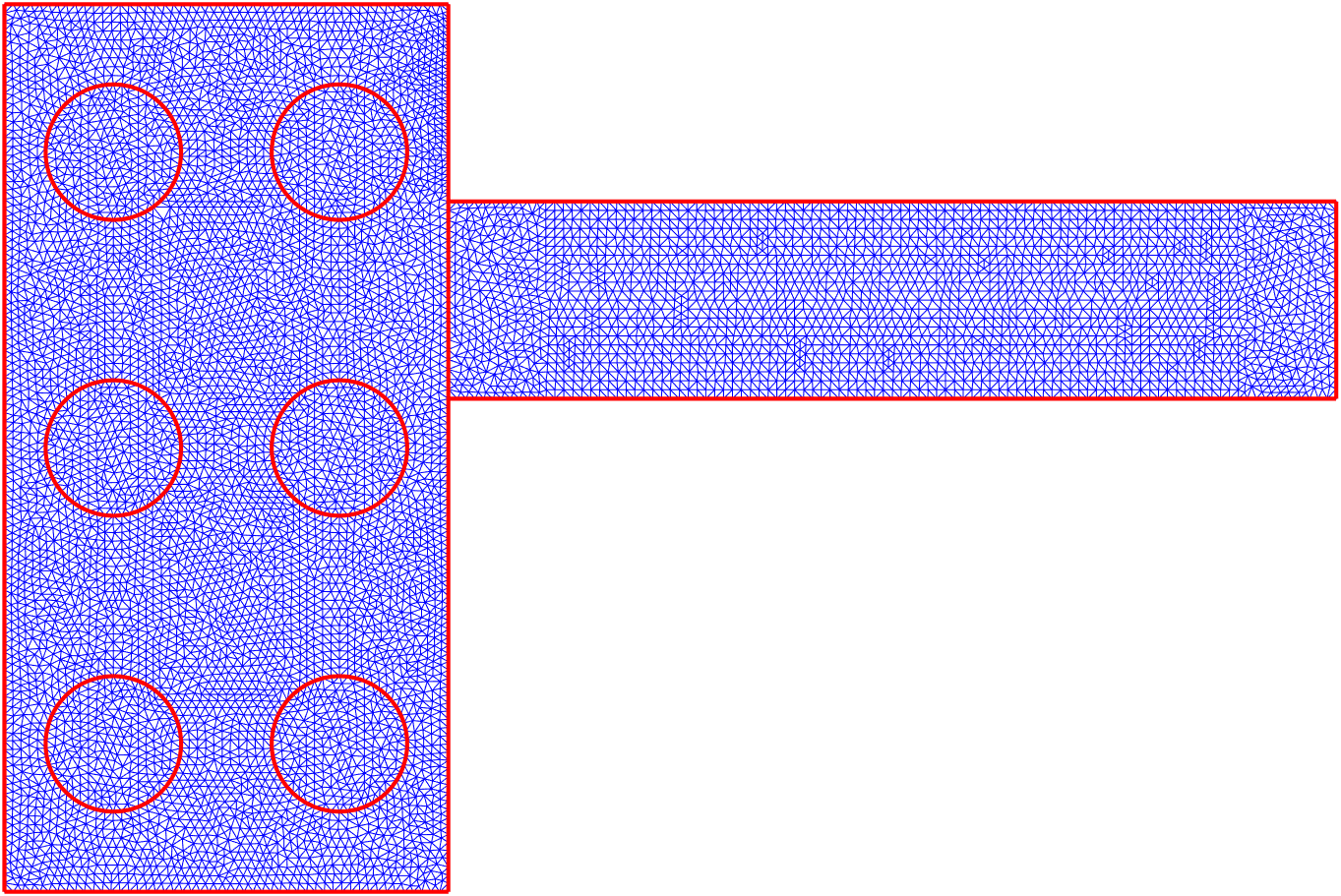}   
    \label{fig:sec6:mesh2}
  \caption{Computational mesh used for computing the
    resonances of the system governed by
    \eqref{eq:sec2:goveq}--\eqref{eq:sec2:bc} in $\Omega$ with $h=4\pi/9$.}
\end{figure}

\begin{figure}[htbp]
  \centering
  \subfigure[Real parts of $\widehat{\lambda}_{6}$ and $\widehat{\lambda}_{7}$.]{
    \centering
        \includegraphics[width=0.9\textwidth]{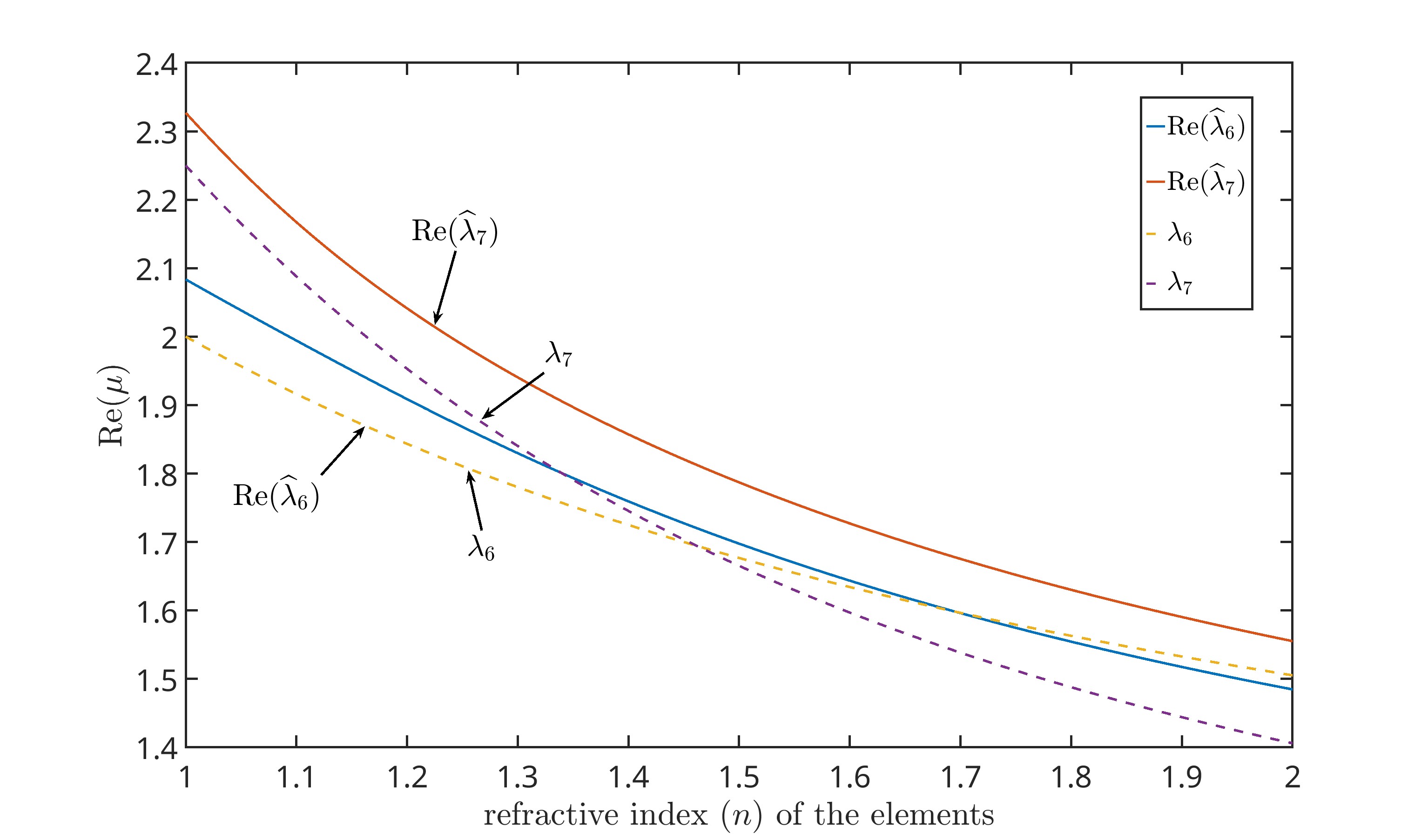}
    \label{fig:sec6:realresonance2}}\\
  \subfigure[Imaginary parts of $\widehat{\lambda}_{6}$ and
  $\widehat{\lambda}_{7}$.]{    
    \centering
        \includegraphics[width=0.9\textwidth]{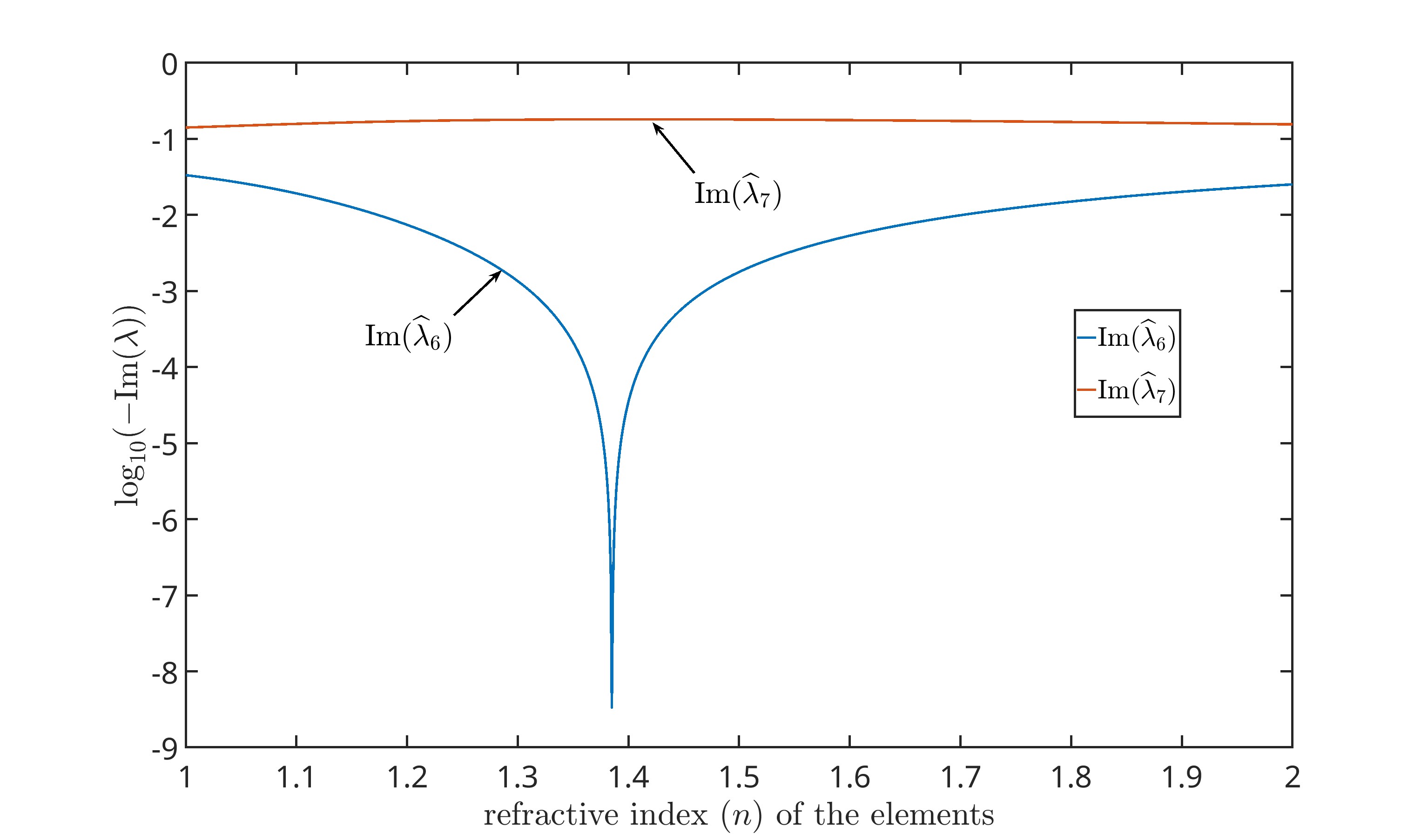}   
        \label{fig:sec6:imagresonance2}}\\
   \subfigure[The BIC mode at $n\approx1.385$ and
      $\mu=\widehat{\lambda}_{6}\approx1.771$.]{        
    \centering
        \includegraphics[width=0.63\textwidth]{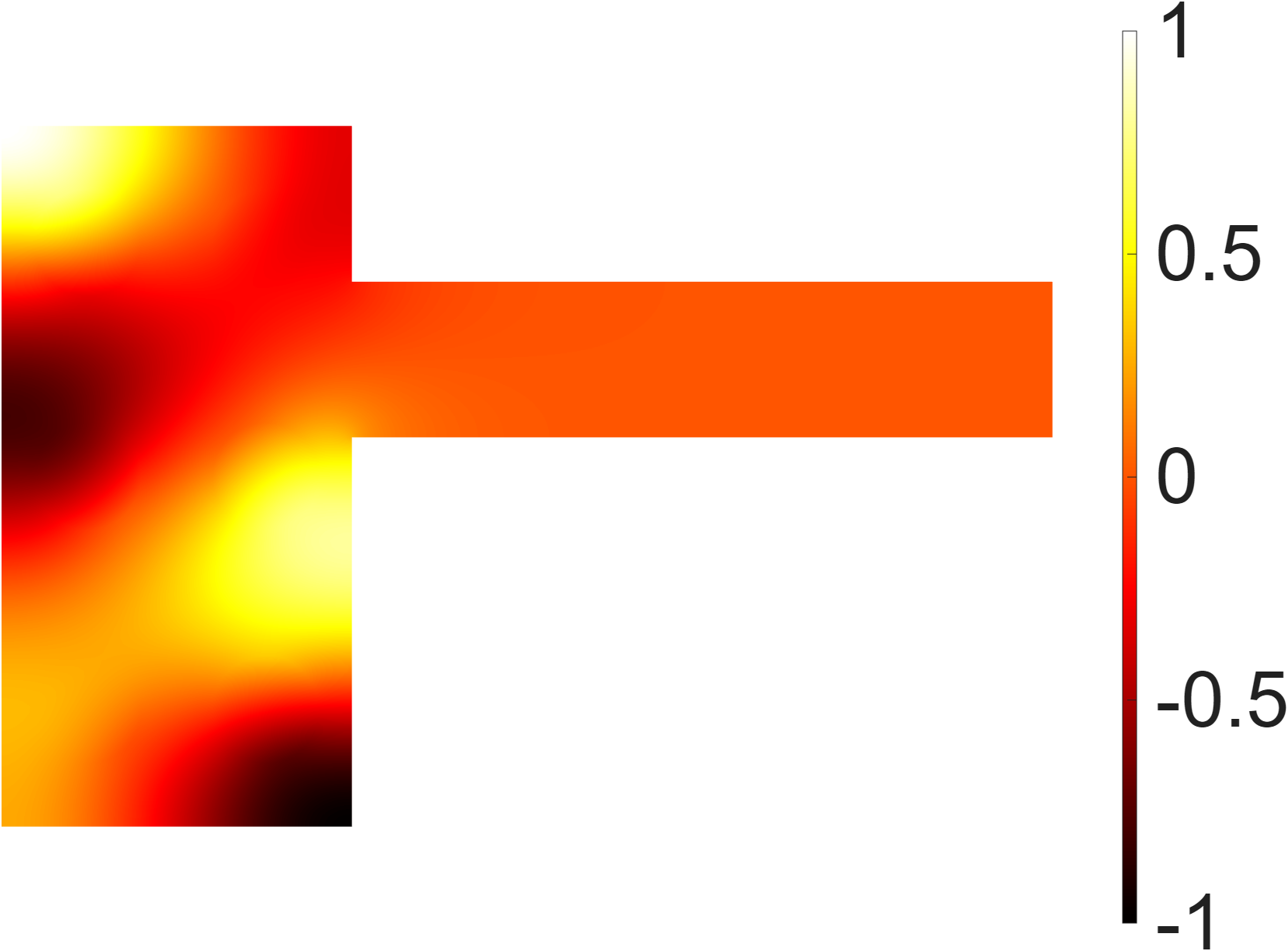}   
    \label{fig:sec6:BIC2}}        
  \caption{Resonances of the problem
    \eqref{eq:sec2:goveq}--\eqref{eq:sec2:bc} in $\Omega$ approximating the
    eigenvalues $\lambda_{6}$ and $\lambda_{7}$ as the refractive index of the
    elements varies. A BIC emerges at $n\approx1.385$ and
      $\mu=\widehat{\lambda}_{6}\approx1.771$.}  
\end{figure}

We further demonstrate the robustness of the
FW-BICs with an example where the waveguide width is increased to
$h=4\pi/9$. The computational mesh for the finite-element simulation is
shown in Fig. \ref{fig:sec6:mesh2}. The resulting resonances
approximating $\lambda_{6}$ and $\lambda_{7}$ are denoted by $\widehat{\lambda}_{6}$ and
$\widehat{\lambda}_{7}$, whose real
and imaginary parts are plotted in Fig. \ref{fig:sec6:realresonance2} and
\ref{fig:sec6:imagresonance2}, respectively. A BIC emerges at $n\approx1.385$ and
$\widetilde{\lambda}_{6}\approx1.771$, and its corresponding mode profile is shown in
Fig. \ref{fig:sec6:BIC2}.

\section{Discussion and conclusion}
In this section, we briefly discuss how to
extend our approach to analyze the existence
of FW-BICs under generalized perturbations, including
parameter-dependent boundary perturbations. Additionally, the
emergence of symmetry-protected BICs in the
cavity-waveguide structures are also investigated.

Previously, we assumed the refractive index $n(\mathbf{x},\delta)\equiv1$ within
$\Omega_{B}$.
If $n(\mathbf{x},\delta)$ also exhibits dependence on $\delta$
in $\Omega_{B}$, our approach remains applicable with minor
modifications. This requires the following conditions:
\begin{itemize}
\item[B.1.] $n(\mathbf{x},\delta)|_{{\overline{\Omega_{B}}}}\in{C^{1,1}(\overline{\Omega_{B}}\times[-\delta_{0},\delta_{0}])}$,  
\item[B.2.] $n(\mathbf{x},\delta)|_{\Gamma_{h}}=C_{\delta}$, where $C_{\delta}$ is a constant
  for each $\delta$.
\end{itemize}
Under such perturbations, the second Green identity 
\eqref{eq:sec3:secondGreen} becomes:
\begin{equation}
  \label{eq:sec7:secondGreen:2}
    (\mathcal{P}u,\psi_{m})_{\Omega_{\mathrm{in}}}-(u,\mathcal{P}\psi_{m})_{\Omega_{\mathrm{in}}}=(u,\frac{1}{n^{2}}\partial_{\bm{\nu}}\psi_{m})_{\partial\Omega_{\mathrm{in}}}-(\frac{1}{n^{2}}\partial_{\bm{\nu}}u,\psi_{m})_{\partial\Omega_{\mathrm{in}}}.
  \end{equation}
Condition B.2 ensures the simplification:
\begin{equation}
  \label{eq:sec7:matching}
  (\frac{1}{n^{2}}\partial_{\bm{\nu}}u,\psi_{m})_{\partial\Omega_{\mathrm{in}}}=\frac{1}{C_{\delta}^{2}}(\partial_{\bm{\nu}}u,\psi_{m})_{\Gamma_{h}}
\end{equation}
by leveraging the boundary condition \eqref{eq:sec2:bc}. Our reduction
techniques then apply directly to the infinite-dimensional
linear system:
\begin{equation}
  \label{eq:sec7:ILS:2}
    \left(\begin{bmatrix}
    \lambda_{0}-\mu&&\\
               &\lambda_{1}-\mu&\\
    &&\ddots
  \end{bmatrix}-\sum_{j=0}^{\infty}\frac{\bi\alpha_{j}}{C_{\delta}^{2}}\mathbf{v}_{j}\mathbf{v}^{*}_{j}\right)
\begin{bmatrix}
  d_{0}\\
  d_{1}\\
  \vdots
  \end{bmatrix}=\mathbf{0}.
\end{equation}
By imposing Condition B.1, we retain the eigenfunction regularity
proven in Lemma \ref{lem:sec5:H3converge}. Consequently, the existence of
FW-BICs analogously follows the analysis presented in Section
\ref{sec:existenceofFWBICs}.

Our approach can also be generalized to
linear boundary perturbations governed by one-to-one transformations:
\begin{equation}
  \label{eq:sec7:bp}
  \bm{\mathcal{T}}_{\delta}:\mathbf{x}\to\mathbf{y}=\mathbf{x}+\delta\bm{\mathcal{R}}(\mathbf{x}).
\end{equation}
The scaling transformation considered in \cite{lyapina15} is
equivalent to defining $\bm{\mathcal{R}}$ as:
\begin{equation}
  \label{eq:sec7:scaling}
  \bm{\mathcal{R}}(\mathbf{x})=(x_{1},0).
\end{equation}
Letting $\Gamma$ denote the line segment on $\partial\Omega_{\mathrm{in}}$ where the
waveguides extend from, we require
\begin{itemize}
  \item[R.1.]  $\bm{\mathcal{R}}(\mathbf{x})$ is a sufficiently smooth function
defined on an open set containing $\overline{\Omega_{\mathrm{in}}}$ and
mapping to $\mathbb{R}^{2}$.
\item[R.2.] The gradient of $\bm{\mathcal{R}}$ satisfies
  \begin{equation}
    \label{eq:sec7:gradientR}
    \nabla\bm{\mathcal{R}}|_{\Gamma}=
    \begin{bmatrix}
      C_{\bm{\mathcal{R}}}&0\\
      0&0
    \end{bmatrix},
  \end{equation}
  where $C_{\bm{\mathcal{R}}}$ is a constant.
\end{itemize}
Under Condition R.2, $\bm{\mathcal{T}}_{\delta}(\Gamma)$ is just a translation of
$\Gamma$. Hence, our perturbed cavity-waveguide region $\Omega_{\delta}$ is
well-defined by assuming the waveguide translates along with $\Gamma$.
For simplicity, we assume $n(\mathbf{x})\equiv1$ in $\Omega$.
Letting $\mathbf{S}_{\delta}:=\nabla(\bm{\mathcal{T}}_{\delta}^{-1})$ and
$J_{\delta}:=\mathrm{det}(\nabla\bm{\mathcal{T}}_{\delta})$, solving the PDE problem
\eqref{eq:sec2:goveq}--\eqref{eq:sec2:bc} in $\Omega_{\delta}$ is equivalent to
solving the following equations in $\Omega$:
\begin{align}
  -\nabla\cdot\left(\mathbf{S}_{\delta}^{T}J_{\delta}\mathbf{S}_{\delta}\nabla{u}\right)+\mu{J_{\delta}}u=&0\
                                                                        \text{in}\ \Omega_{\mathrm{in}},\label{eq:sec7:omegin:goveq}\\
  \bm{\nu}\cdot\mathbf{S}_{\delta}^{T}J_{\delta}\mathbf{S}_{\delta}{\nabla}u=&0\ \text{on}\ \partial\Omega_{\mathrm{in}}/\Gamma_{h},\label{eq:sec7:omegin:bc}\\
  -\Delta{u}+\mu{u}=&0\ \text{in}\
               \Omega_{\mathrm{out}},\label{eq:sec7:omegout:goveq}\\
  \partial_{\bm{\nu}}u=&0\ \text{on}\ \partial\Omega_{\mathrm{out}}/\Gamma_{h},\label{eq:sec7:omegout:bc}\\  
  u|_{\Gamma_{h,-}}=&u|_{\Gamma_{h,+}},\label{eq:sec7:gammah:trace}\\
  \bm{\nu}\cdot\mathbf{S}_{\delta}^{T}J_{\delta}\mathbf{S}_{\delta}{\nabla}u|_{\Gamma_{h,-}}=&\partial_{x_{1}}u|_{\Gamma_{h,+}},\label{eq:sec7:gammah:conormal}                 
\end{align}
where $\Gamma_{h,-}$ and $\Gamma_{h,+}$ denote the sides of $\Gamma_{h}$ approached from
$\Omega_{\mathrm{in}}$ and $\Omega_{\mathrm{out}}$ respectively.
In the $\Omega_{\mathrm{in}}$ region, the eigenfunction expansion remains valid
by considering the following generalized eigenvalue problem: 
\begin{align}
    -\nabla\cdot\left(\mathbf{S}_{\delta}^{T}J_{\delta}\mathbf{S}_{\delta}\nabla{\psi}\right)+\lambda{J_{\delta}}\psi=&0\
                                                                        \text{in}\ \Omega_{\mathrm{in}},\label{eq:sec7:omegin:eigen:goveq}\\
  \bm{\nu}\cdot\mathbf{S}_{\delta}^{T}J_{\delta}\mathbf{S}_{\delta}{\nabla}\psi=&0\ \text{on}\ \partial\Omega_{\mathrm{in}},\label{eq:sec7:omegin:eigen:bc}
\end{align}
where the eigenfunctions form a complete orthonormal basis in
$L^{2}(\Omega_{\mathrm{in}})$ with the weight $J_{\delta}$. The continuity of
the eigenvalues and eigenfunctions as $\delta$ varies can also be
established
under the transformation \eqref{eq:sec7:bp} and Condition R.1.
(cf. Section VII.6.5 in \cite{kato95}). By applying a similar
mode-matching method and invoking Condition R.2, we derive the
following infinite-dimensional linear system:
\begin{equation}
  \label{eq:sec7:ILS:3}
    \left(\begin{bmatrix}
    \lambda_{0}-\mu&&\\
               &\lambda_{1}-\mu&\\
    &&\ddots
  \end{bmatrix}-\sum_{j=0}^{\infty}\frac{\bi\alpha_{j}}{1+\delta{C}_{\bm{\mathcal{R}}}}\mathbf{v}_{j}\mathbf{v}^{*}_{j}\right)
\begin{bmatrix}
  d_{0}\\
  d_{1}\\
  \vdots
  \end{bmatrix}=\mathbf{0}.
\end{equation}
By requiring that Condition A.2 and A.3 are satisfied, the existence of
FW-BICs can be established via an argument analogous to the analysis
presented in Section \ref{sec:existenceofFWBICs}.

Symmetry-protected BICs are a class of BICs frequently observed
in symmetric
structures. Their existence arises from a symmetry mismatch between
the bound states and the
outgoing radiating waves \cite{anne94,evans94,shipman07}.
The mode-matching method can be used to rigorously establish the
existence of
symmetry-protected BICs in cavity-waveguide
structures, if the waveguide widths are sufficiently small.
Assuming
the region $\Omega$ is symmetric with
respect to $x_{2}=0$, the system
\eqref{eq:sec3:ILS}--\eqref{eq:sec3:vn} decouples into two subsystems
governed by eigenfunctions with the following symmetries:
\begin{itemize}
  \item even: $\psi^{(e)}(x_{1},x_{2})=\psi^{(e)}(x_{1},-x_{2})$,
  \item odd: $\psi^{(o)}(x_{1},x_{2})=-\psi^{(o)}(x_{1},-x_{2})$.
  \end{itemize}
For the odd subsystem,
\eqref{eq:sec4:ILS:imag} is automatically satisfied. Following the
reduction techniques outlined in Section \ref{sec:reduction}, we obtain
the governing equation:
\begin{equation}
  \label{eq:sec7:spBIC}
  \lambda-\mu+a(\mu,h)=0,
\end{equation}
where $\lambda$ is a simple eigenvalue associated with an odd
eigenfunction and $a(\mu,h)$ represents a function depending on $\mu$ and
$h$. By
demonstrating the uniform convergence of $a(\cdot,h)$ to $0$
as $h\to0$ and the continuity of $a(\mu,h)$ in $\mu$, the existence of a BIC is
guaranteed for $h\ll1$ by the Intermediate Value Theorem.

\bibliographystyle{siamplain}
\bibliography{references}
\end{document}